\documentclass[twocolumn,pra,aps, showpacs]{revtex4-1}
\usepackage{xspace,amsmath,amsfonts,amsthm,amssymb,amsbsy,graphics,graphicx,bm,epstopdf}
\usepackage[usenames,dvipsnames]{color}
\usepackage{soul} 
\usepackage[breaklinks=true]{hyperref}
\hypersetup{
  colorlinks   = true, 
  urlcolor     = blue, 
  linkcolor    = blue, 
  citecolor   = red 
}
\newcommand{\smallfrac}[2]{\mbox{$\frac{#1}{#2}$}}
\newcommand{\half}{\smallfrac{1}{2}}
\newcommand{\quart}{\smallfrac{1}{4}}

\newcommand{\bra}[1]{\left\langle{#1}\right|}
\newcommand{\ket}[1]{\left|{#1}\right\rangle}
\newcommand{\op}[2]{\ket{#1}\!\bra{#2}}

\newcommand{\D}[1]{\sD\left[{#1}\right]}
\newcommand{\sB}{\mathcal{B}}

\newcommand{\sD}{\mathcal{D}}
\newcommand{\sE}{\mathcal{E}}
\newcommand{\sF}{\mathcal{F}}
\newcommand{\sG}{\mathcal{G}}
\newcommand{\sH}{\mathcal{H}}

\newcommand{\sS}{\mathcal{S}}
\newcommand{\sU}{\mathcal{U}}

\newcommand{\Rt}{R_{\mathrm{th}}}
\newcommand{\Pt}{P_{\mathrm{th}}}
\newcommand{\Pbart}{\bar P_{\mathrm{th}}}
\newcommand{\Ptvalue}{0.135} 
\newcommand{\Pbartvalue}{0.133} 
\newcommand{\tauzerovalue}{7.89} 
\newcommand{\tauonevalue}{0.722} 
\newcommand{\taubarvalue}{0.718} %

\newcommand{\data}{D}
\newcommand{\Hzero}{\sH_\emptyset}
\newcommand{\Hone}{\sH_1}
\newcommand{\tr}{\mbox{\rm tr}}
\newcommand{\erf}[1]{Eq.~(\ref{#1})}
\newcommand{\frf}[1]{Fig.~\ref{#1}}
\newcommand{\srf}[1]{Sec.~\ref{#1}}
\newcommand{\arf}[1]{App.~\ref{#1}}
\newcommand\dg{^\dagger}

\definecolor{orange}{rgb}{1,0.5,0}
\setstcolor{Magenta}

\newcommand\prob{\mbox{Pr}}

\begin{document}
\title{In-situ characterization of quantum devices with error correction}

\author{Joshua Combes$^1$}\email{joshua.combes@gmail.com}
\author{Christopher Ferrie$^1$}
\author{Chris Cesare$^1$}
\author{\\Markus Tiersch$^{2,3}$}
\author{G. J.~Milburn$^4$}
\author{Hans J.~Briegel$^{2,3}$}
\author{Carlton M.~Caves$^{1,4}$}\email{ccaves@unm.edu}

\affiliation{$^1$Center for Quantum Information and Control, University of New Mexico, Albuquerque, New Mexico, 87131-0001\\
$^2$Institut f{\"u}r Theoretische Physik, Universit{\"a}t Innsbruck, Technikerstra{\ss}e 25, A-6020 Innsbruck\\
$^3$Institut f{\"u}r Quantenoptik und Quanteninformation der \"Osterreichischen Akademie der Wissenschaften, Innsbruck, Austria\\
$^4$Centre for Engineered Quantum Systems, School of Mathematics and Physics, The University of Queensland, St Lucia, QLD 4072, Australia}

\pacs{03.67.Pp, 03.67.Ac, 06.20.-f, 02.50.-r}

\begin{abstract}
Syndrome measurements made in quantum error correction contain more information than is typically used.  We show that the statistics of data from syndrome measurements can be used to do the following: (i)~estimation of parameters of an error channel, including the ability correct away the invertible part of the error channel, once it is estimated; (ii)~hypothesis testing (or model selection) to distinguish error channels, e.g., to determine if the errors are correlated.  The unifying theme is to make use of all of the information in the statistics of the data collected from syndrome measurements using machine learning and control algorithms.
\end{abstract}

 \date{\today}

\maketitle

\section{Introduction}\label{sec:intro}
Quantum error correction is one of the most surprising and important developments in quantum theory.  Without error correction and related techniques of fault tolerance there is no justification to believe that a universal quantum computer is possible in practice.   Since the early days of the Shor~\cite{Shor95} and Steane~\cite{Steane96} codes, quantum error correction has seen many developments, notably Gottesman's stabilizer formalism~\cite{Got96} and Kitaev's topological codes~\cite{Kitaev97}. In all of these, a quantum state is encoded into an error-correcting code, which can protect the quantum state against the high-probability errors induced by an error channel. Failure arises from low-probability errors that cannot be detected and corrected by a particular code in a single shot. The cost exacted by error correction is an increase in the number of qubits, gates, and measurements required to perform the computation.


In practice, a scalable quantum computer will be a highly integrated system. Highly integrated implementations of large-scale quantum computing will likely be fabricated all at once, so all the layers above the physical qubits, i.e., error correction, logical operations, device characterization, and calibration operations, must be built in from the beginning.  It is often imagined that modular single- and two-qubit gates can be constructed, characterized, and then composed without introducing additional error, but in a highly integrated implementation, one must face the question of how to integrate calibration and characterization with the functional parts of the system.

In this paper we provide a partial answer by showing how the statistics of error-correction syndrome measurements can be used to perform in-situ characterization, i.e., to estimate or to distinguish between error channels.  For example it is possible to describe an imperfect unitary gate as a perfect gate followed by (or preceded by) imperfections that take the encoded qubits out of the logical subspace. Our method allows one to estimate these imperfections and thus to characterize some of the gate imperfection.  Specifically, we show the following:
\begin{itemize}
\item[1.~]Using  parameter estimation, one can learn about different parameters, unitary and random, of an error channel, provided that the parameters make distinguishable contributions to the syndrome probabilities.  Once estimated, the unitary part of an error channel can be corrected away.
\item[2.~]Using techniques of hypothesis testing or model selection, one can distinguish error-channel models that give different syndrome probabilities.
\item[3.~]Control mechanisms can change and thus differentiate the otherwise degenerate contributions that different parameters or models make to the syndrome probabilities; i.e., control can be used to make estimation and discrimination easier.
\end{itemize}

We stress that the estimation or hypothesis testing is done on a single encoded system; the error-correction protocol re-initializes the system in the code subspace after each round of error correction.  Importantly, we can do the parameter estimation or hypothesis testing even if the encoded quantum information is completely destroyed, because these work identically for any state, pure or mixed, in the code subspace.  In situations where the errors are small enough, however, the estimation or hypothesis testing can be performed while protecting the quantum information encoded in the code subspace.

For simplicity we illustrate the general ideas with examples that use the three-qubit repetition (bit-flip) code.   In \srf{sec:Param_est} we show that in an error channel that contains systematic (unitary) errors and random (incoherent or decohering) errors, both types of errors can be estimated by using control to lift the degeneracy between the contributions of the systematic and random errors to the syndrome probabilities.  This illustrates of points~1 and~3 above.  It is especially interesting because an agent, after learning the unitary part of the channel, could permanently correct away this unitary part, thereby reducing the frequency of error-correction rounds.  Section~\ref{sec:Hypothesis_test} turns to hypothesis testing and model selection.  We demonstrate that two error-channel models, one having uncorrelated errors on the physical qubits and the other having correlated errors, can be distinguished by using the statistics of syndrome measurements, thus illustrating point~2 above.

The numerical methods that we use here are examples of machine learning and control~\cite{MCMC}.  We anticipate scenarios where the data collection and control processes become more autonomous and adaptive so that they can adapt to changes in the environment, e.g., to a changing noise channel or to an increasing temperature in the environment that gives rise to an increasing error rate.  Sufficiently elaborated, such autonomous and adaptive processes can be seen as instances of intelligent agents~\cite{BriDelas12} that process and control quantum systems.

We sketch, in Sec.~\ref{sec:therealthing}, how the ideas of \srf{sec:Param_est} could be applied to any stabilizer code, working out as an example the five-qubit ``perfect'' quantum code that corrects all single-qubit errors~\cite{LafMiqPazZur96,BenDiVSmoWoo96}.

In \srf{sec:prior} we consider related work on applications of error-correcting codes to metrology and other estimation tasks.  In particular, we comment on a recent series of papers~\cite{DurSkoFroKra13,Ozeri13,KesLovSusLuk13,ArrVinAhaRet13} that proposed using error-correcting codes for quantum metrology in lossy, decoherent systems.  In addition, during the final stages of preparation of this manuscript, two papers that study using syndrome-measurement outcomes to perform estimation tasks were posted to the arXiv e-print server~\cite{FowSankKelBarMar2014,OmSriBan2014}; we compare and contrast these proposals with our methods in \srf{sec:prior}.

We conclude in \srf{sec:discussion} with a discussion of extensions of our work and open questions.

\begin{figure*}[ht!]
\includegraphics[width=0.85\linewidth]{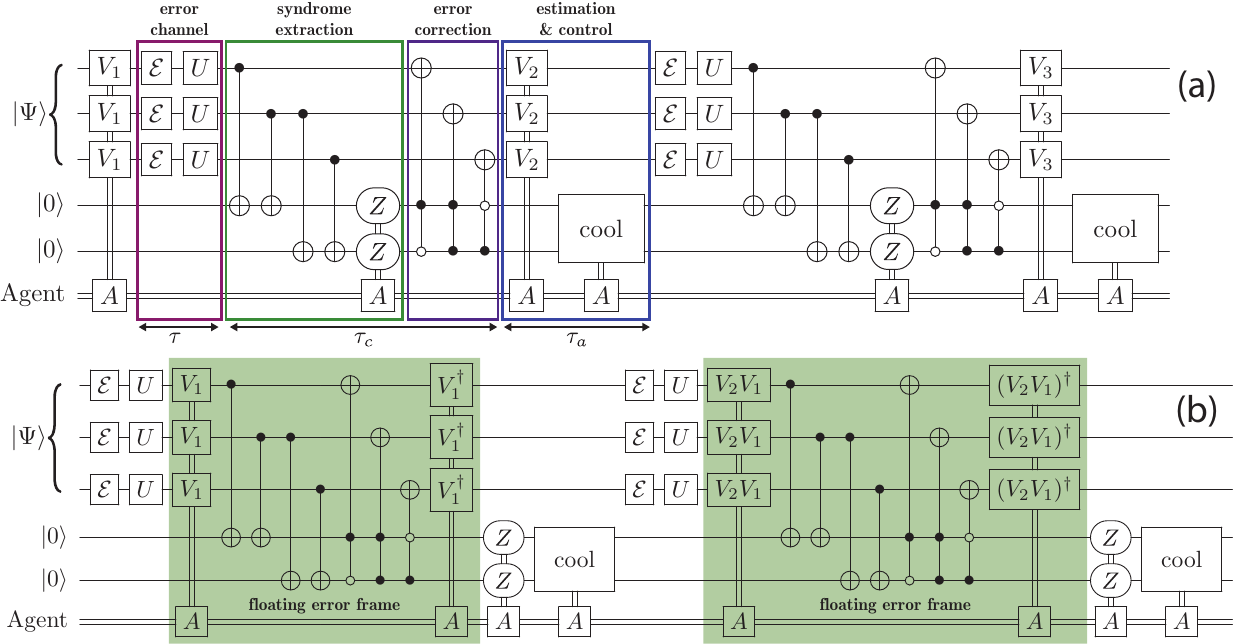}\\
\caption{\label{fig1}
Two rounds of a protocol for estimating and correcting away a systematic local unitary $U\equiv\exp(-i\omega\tau X/2)$, which rotates each qubit by angle $\omega\tau$ about the Bloch $x$ axis.  The qubit state $\ket{\psi}=\alpha\ket{0}+\beta\ket{1}$ is encoded into the three-qubit bit-flip repetition code as the state $\ket{\Psi} = \alpha \ket{000} +\beta\ket{111}$.  The error channel acts independently on each of the three qubits and is a composition, on each qubit, of the unitary map $U$ and the nonunitary map $\sE(\rho)=(1-p)\rho+pX\rho X$, which bit-flips each qubit with probability $p$. (a) Based on prior information about the unitary part of the channel, the agent applies a control unitary $V_c^{\otimes 3}$ (labeled by round number in the circuit diagram), where $V_c=\exp(i\omega_c\tau X/2)$, to counteract the effect of $U$.  This counter-unitary could be applied continuously throughout the time the error channel acts; this would continuously combat the persistent rotation $U$, but depicting such continuous control is not easy in a circuit diagram, so the counter-unitary is shown as temporally distinct in the diagram.  The usual ancilla-coupled parity syndrome measurements and subsequent controlled-unitary corrections are performed.  In this rendering, the correction unitaries are depicted coherently, but since the agent gets the syndrome data by measuring the ancillas, the correction operations could be classically conditioned on the agent's syndrome information.  The agent combines the syndrome data with prior information to estimate the channel parameters and uses the estimates to plan the next round of counter-rotations.  The agent also uses the syndrome data to reprepare the ancilla qubits in the state $\ket0$ (alternatively, new ancilla qubits could be swapped in), a task symbolized by the box labeled ``cool''; the entropy introduced by the error channel is left in the agent.  The circuit diagram indicates the duration $\tau$ over which the error channel acts in each round, the time $\tau_c$ required for syndrome detection and error correction, and the time $\tau_a$ required by the agent to apply counter-unitaries and to cool the ancilla qubits.  Throughout our analysis, we assume that $\tau_c$ and $\tau_a$ are so small compared to $\tau$ that they can be neglected, and we refer to $\tau$ as the duration of an error-correction round.  (b) Instead of correcting the external field using the control unitary $V_c$, the countering effects of the control unitary can be incorporated by rotating the Pauli basis of the syndrome and error-correcting operations on the three physical qubits, i.e., using Pauli operators $\tilde X=X$, $\tilde Y=V_c^\dag YV_c$ and $\tilde Z=V_c^\dag Z V_c$.  In the circuit diagram, this ``floating Pauli frame'' is depicted by the shaded regions, in which the bracketing control unitaries can be regarded as conjugating the Pauli basis of the physical qubits for the intervening operations.  Notice that the measurements that make the syndrome data available to the agent, plus the processing of the syndrome data by the agent, the cooling operation, and the floating-frame rotations, could all be done reversibly and coherently, in which case the agent could be regarded as a ``quantum agent.''  Such a quantum agent would eventually have to come into contact with the external environment, in order to erase its memory in preparation for further rounds of the protocol.}
\end{figure*}

\section{Parameter estimation}\label{sec:Param_est}

Consider an arbitrary qubit state $\ket{\psi}=\alpha\ket{0}+\beta\ket{1}$, where $\ket0$ and $\ket1$, the standard basis states, are eigenstates of the Pauli operator $Z=\op{0}{0}-\op{1}{1}$.  In the three-bit repetition code, $\ket\psi$ is encoded as the logical state $\ket{\Psi} = \alpha\ket{000} +\beta\ket{111}$, which lies in the code subspace spanned by $\ket{000}$ and $\ket{111}$. The goal of this section is to demonstrate that it is possible to estimate error-channel parameters from the statistics of the syndrome measurements.

Before describing the error-correction process, we need to describe the error channel. In this section we assume that we have an error channel $\sF$ described by the composition of a nonunitary process $\sE^{\otimes3}$ and a unitary process $\sU^{\otimes3}$, i.e., $\sF=\sU^{\otimes3}\circ\sE^{\otimes3}$.  As is illustrated in \frf{fig1}(a), both the unitary and nonunitary processes act independently and identically on the three physical qubits.

The unitary part of the channel is $\sU(\rho)=U\rho\,U^\dagger$, where
\begin{equation}
U\equiv e^{-i\omega\tau X/2}=I\cos(\omega\tau/2)-iX\sin(\omega\tau/2)
\end{equation}
rotates a qubit by angle $\omega\tau$ about the $x$ axis of the Bloch sphere, $\omega$ being a frequency, ${\tau}$ the time interval over which the error channel acts, and $X=\op{0}{1}+\op{1}{0}$ the bit-flip Pauli operator.  Such a persistent unitary channel, acting identically on the three physical qubits, might arise, for example, from a stray magnetic field.  We assume that the rotation axis, here the $x$ axis, is fixed and that $\omega$ is unknown and to be estimated; the restriction to a single rotation axis is lifted in \srf{sec:therealthing}.

The nonunitary part of the channel is described by a quantum operation~\cite{MikeandIke} that flips a qubit with probability $p$, i.e.,
\begin{equation}\label{eq:Epx}
\sE(\rho)=(1-p)\rho + pX\rho X.
\end{equation}
Since one of the control mechanisms we use in the following is to vary the error channel's duration $\tau$, we need to model how the nonunitary part of the channel depends on $\tau$, something done automatically for the unitary part.

The natural model uses an open quantum systems approach to model the entire channel.  Specifically we use the vacuum master equation $d\rho = -i\,dt\,[H,\rho] +2\gamma\,dt\,\D{c}\rho$, where $H$ is a Hamiltonian, $\gamma$ is a rate that describes how strongly the system and bath are coupled, $c$ is an arbitrary operator on the system, and the superoperator $\D{A}B$ is defined as $\D{A}B\equiv A B A\dg -\half(A\dg AB+BA\dg A)$.  To model the single-qubit channel $\sU\circ\sE$ we take $c= X$ and $H=\half\omega X$, which gives the master equation
\begin{align}\label{eq:vacuum_master}
d\rho &= -i\smallfrac{1}{2}\omega\,dt\,[X,\rho] + 2\gamma\,dt\,(X\rho X - \rho),
\end{align}
describing damped Rabi oscillations. Because the Hamiltonian and diffusion parts of the master equation commute, i.e. $U\sE(\rho)U\dg=\sE(U\rho U\dg)$, the master equation is easy to solve, and its solution yields the time dependence of $p$ on the time interval $\tau$~\cite{pnote}:
\begin{align}\label{p_timedepend}
p= \half (1- e^{-4\gamma \tau}).
\end{align}
Notice that for all $\gamma\tau$, $p\in[0,1/2]$.  The task of characterizing the random errors is the job of estimating~$\gamma$.

A nice way of expressing the combined action of $\sE$ and $\sU$ leverages the fact that in stabilizer codes syndrome measurements project all errors to Pauli errors. It is possible to derive an effective single-qubit error rate $P$ and to ignore the coherences introduced by $U$. The combined action of the single-bit channel over a time interval $\tau$ can be expressed as
\begin{align}\label{singlebitchannel}
\sU\circ\sE(\rho)&=Q\rho+PX\rho X-iC[X,\rho].
\end{align}
Here
\begin{align}\label{eq:P}
P&=\sin^2(\omega\tau/2)+p\cos\omega\tau\nonumber\\
&=\smallfrac{1}{2}(1-e^{-4\gamma\tau }\!\cos\omega\tau)
\simeq\smallfrac14(\omega\tau)^2+2\gamma\tau
\end{align}
is the probability of a bit flip, including both the unitary and nonunitary contributions.   The final form in \erf{eq:P} gives the dominant contributions when $(\omega\tau)^2,\gamma\tau\ll1$; in this small-$\tau$ approximation, which is the same as assuming small error probability, $\omega$ and $\gamma$ make separate contributions to $P$.  The quantity $Q=1-P$ is the probability not to flip,
\begin{align}\label{eq:Q}
Q=\cos^2(\omega\tau/2)-p\cos\omega\tau
=\smallfrac{1}{2}(1+e^{-4\gamma\tau}\!\cos\omega\tau),
\end{align}
and
\begin{equation}\label{eq:C}
C=\smallfrac12(1-2p)\sin\omega\tau=\smallfrac12e^{-4\gamma\tau}\sin\omega\tau
\end{equation}
describes the development and decay of coherences in the standard basis.

An error-correction round consists of syndrome extraction followed by conditional unitaries that correct the high-probability errors, as shown in \frf{fig1}(a).  Bit-flip errors are detected by measurements of the stabilizer generators of the repetition code, here taken to be measurements of $Z\otimes Z\otimes I$ and $I\otimes Z\otimes Z$.  The first of these checks the parity of the first two qubits, and the second checks the parity of the second two qubits.  The four possible results of these syndrome measurements are labeled by 00, 10, 01, and 11, where a zero denotes even parity and a 1 denotes odd parity.  The syndrome measurements discretize the errors, forcing the system to commit to bit-flip errors.  The quantity $C$ of \erf{eq:C} thus becomes irrelevant to the syndrome measurements, and the probabilities of bit flips on the three qubits are described by $P$ and $Q$.  The probability for $m$ bit flips on the three qubits is
\begin{equation}\label{err_prob1}
p^e_m=\frac{3!}{m!(3-m)!}P^mQ^{3-m}.
\end{equation}

The syndrome 00 corresponds to no bit flips, in which case the system remains in the state $\ket\Psi$, or to bit flips on all three qubits, in which case the system is left in the state $\ket{\Psi'}=(\alpha\ket{111}+\beta\ket{000})/\sqrt2$.  Consequently, the probability for the 00 syndrome is $\prob(00)=Q^3+P^3$.  The syndrome 10 corresponds to a bit flip on the first qubit, in which case the system is left in the state $(\alpha\ket{100}+\beta\ket{011})/\sqrt2$, or to bit flips on the second two qubits, in which case the system is left in the state $(\alpha\ket{011}+\beta\ket{100})/\sqrt2$; thus the probability for this syndrome is $\prob(01)=PQ^2+P^2Q=PQ$.  Similarly, the syndrome 01 corresponds to a bit flip on the third qubit or to bit flips on the first two qubits, and the syndrome 11 corresponds to a bit flip on the second qubit or to bit flips on the first and third qubits.  The syndromes 01 and 11 each have the same probability as the syndrome 10.

The syndrome probabilities, given the channel parameters $\omega$ and $\gamma$, i.e., the likelihood functions, are thus given~by
\begin{subequations}\label{likelihood}
\begin{align}
\prob(00|\omega,\gamma)&=P^3+Q^3=\quart(1+3 e^{-8\gamma\tau}\cos^2\omega\tau).\\
\prob(S|\omega,\gamma)&=PQ=\quart(1-e^{-8\gamma\tau}\cos^2\omega\tau).
\end{align}
\end{subequations}
where $S\in\{10,01,11\}$.  The syndrome probabilities~(\ref{likelihood}) do not depend on $\alpha$ or $\beta$; this is generically true for any error-correcting code because otherwise the syndrome measurements would reveal information about the encoded state.  An important implication, especially important for applications to quantum sensors and quantum metrology, is that all initial states, pure or mixed, in the code subspace have the same efficacy for estimating the channel parameters.  Thus our protocol can be used to characterize channels without assuming any state preparation other than preparation within the code subspace.

For the cases of no bit flips or one bit flip, the post-syndrome states are in one of four orthogonal subspaces, which, given the syndrome, can be mapped unitarily back to the code subspace, thus correcting the error and restoring the logical state $\ket\Psi$.  For the cases of two or three bit flips, the post-syndrome states are in the same orthogonal subspaces, but the conditional error-correction unitaries map the system to $\ket{\Psi'}$ instead of $\ket\Psi$.  This does not affect the ability of further error-correction rounds to determine the channel parameters, but it does mean that the quantum information encoded in the logical state $\ket\Psi$ is lost.  The total probability of uncorrectable errors is
\begin{align}\label{eq:R}
R=p^e_2+p^e_3=3P^2Q+P^3\simeq3P^2.
\end{align}
The final form is the dominant contribution when $P\ll1$, in which case single-qubit flips are the dominant error; running for roughly $1/R\simeq1/3P^2$ error-correction rounds leads to a sizeable probability for encountering an uncorrectable error.

When $\alpha=\beta=1$ or $\alpha=-\beta=1$, the encoded state is a fixed point of the error correction, returned for all numbers of single-qubit errors.  More generally, the overlap of $\ket\Psi$ and $\ket{\Psi'}$ and, hence, the fidelity after uncorrectable errors depends on $\alpha$ and $\beta$; in addition, two uncorrectable errors returns the state to the initial state.  These facts, which do not hold for error-correcting codes that correct all single-qubit errors, make it problematical to use fidelity as the measure of losing the encoded quantum information, so in what follows, we use instead the probability of no uncorrectable errors after $N$ rounds of error correction.

As noted above, the syndrome measurements discretize the errors and commit the system to bit flips.  The entropy introduced by the errors is channeled into the ancilla qubits that are used to make the syndrome measurements and thence, in the depiction of \frf{fig1}, into the agent that runs the error-correction rounds and controls the system based on what it learns from the syndrome measurements.  For the nonunitary part of the channel, the entropy increase is inevitable.  The unitary part, however, coherently rotates the encoded state out of the code subspace, leaving the entropy unchanged; it is the syndrome measurements that discretize this rotation and commit it to a digital error, thus increasing the entropy.  If we know how the code subspace rotates, we can undo the rotation by the controls of \frf{fig1}(a) or take the rotation into account by adjusting the Pauli frame as in \frf{fig1}(b).

We turn now to how to use the syndrome data to estimate both the unitary ($\omega$) and nonunitary ($\gamma$) parts of the channel, thus allowing us to correct away the unitary rotation.  Standard Bayesian or frequentist estimation techniques can be applied to the syndrome likelihoods~(\ref{likelihood}).  There is, however, an immediate problem: the likelihoods are functions only of $P$ (and $Q=1-P$) and thus have a degeneracy in $\gamma$ and $\omega$; if one performs many identical rounds of error correction, one can estimate $P$ with increasing accuracy from the accumulated syndrome data, but cannot distinguish values of $\gamma$ and $\omega$ that give the same value of $P$.  Our key insight is that the unitary and nonunitary parts of the channel can be distinguished by any means that breaks their degeneracy in the likelihood functions.

\begin{figure}[ht!]
\includegraphics[width=0.95\linewidth]{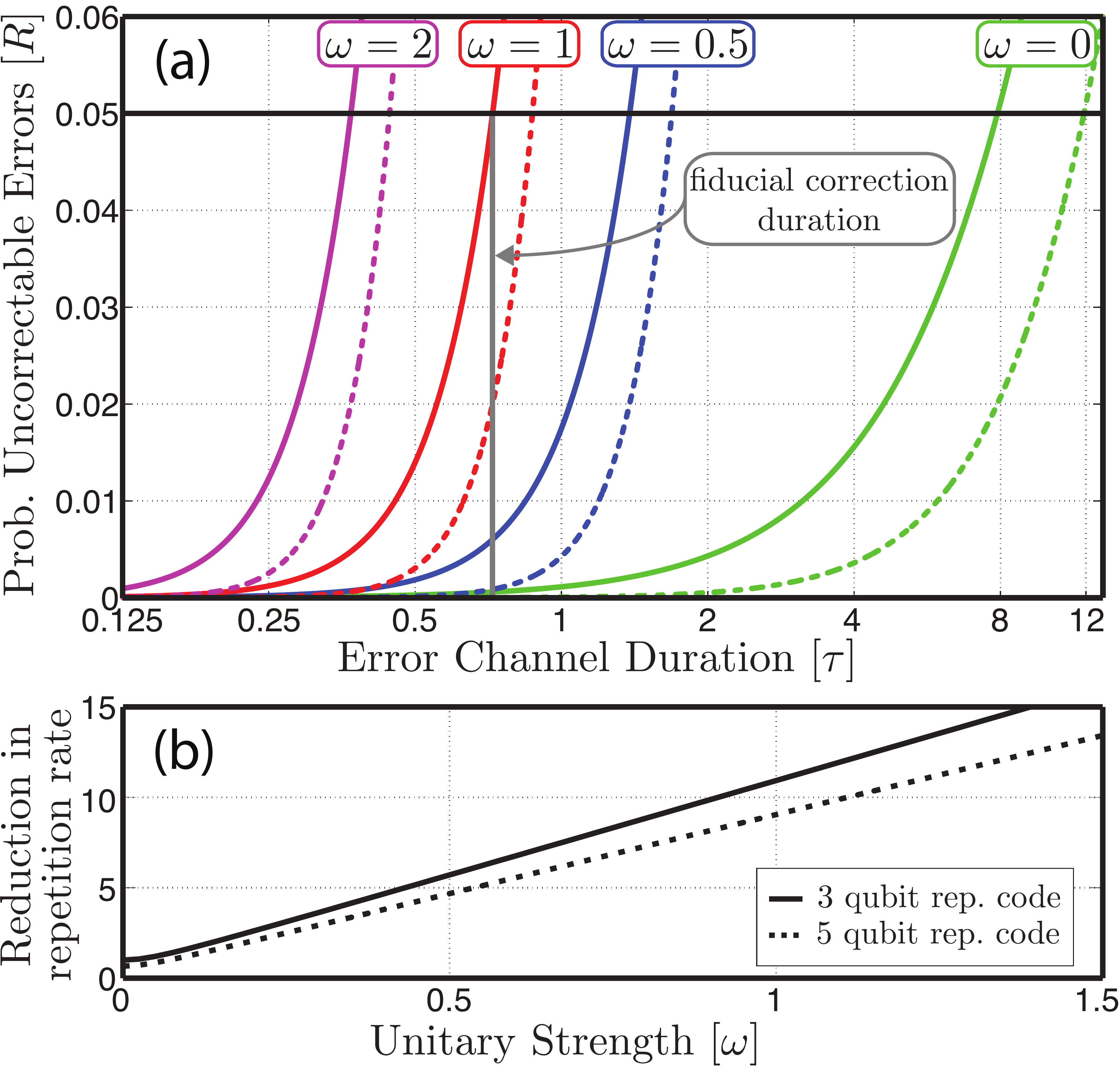}
\caption{\label{fig2}
Exploration of repetition rate for error-correction rounds.  (a)~Solid lines give the probability per round of uncorrectable errors, $R=p_e(2)+p_e(3)$, for the three-qubit repetition code with $\gamma=0.01$ and for four values of $\omega$: $\omega=0$ (green), $\omega=0.5$ (blue), $\omega=1$ (red), and $\omega=2$ (purple).  For a given $\omega$, the duration $\tau$ at which $R$ exceeds the threshold value $\Rt=0.05$ ($\Pt=\Ptvalue$) we call $\tau_\omega$.  When $\omega=1$, the error correction must be performed at $\tau\le\tau_1=\tauonevalue$ (grey line, labeled by ``fiducial correction duration'') to keep the probability of uncorrectable errors less than or equal to our threshold.  If $\omega=0$, error correction must be performed at times $\tau\le\tau_0=\tauzerovalue$ to keep within our threshold.  The dashed lines give the probability per round of the uncorrectable errors for the five-qubit repetition code, for which $R=p^e_3+p^e_4+p^e_5=10P^3Q^2+5P^4Q+P^5$; this illustrates that a higher distance code has a lower $R$ for the same channel parameters.  (b)~Factor $\tau_0/\tau_\omega$ by which the required error-correction repetition rate is reduced, as a function of $\omega$, when the systematic rotation is countered by an exactly counteracting controlling unitary $V_c$, i.e., $\omega-\omega_c=0$.  The factor can also be thought of as the increase in repetition rate, relative to $\omega=0$ (for the three -qubit repetition code), required to keep $R$ below the threshold in the absence of any counter-unitary.  The dashed lines are for the five-qubit repetition code.}
\end{figure}

We explore here two techniques for breaking the degeneracy and thus estimating $\omega$ and $\gamma$ separately.  The first, depicted in \frf{fig1}, is a controlled counter-rotation about $X$, which changes the values of the trigonometric functions in the likelihoods~(\ref{likelihood}).  In particular, the control unitary $V_c=\exp(i\omega_c t X/2)$ applied to the three physical qubits modifies the likelihoods~(\ref{likelihood}) so that $\omega\mapsto\omega-\omega_c$; a counter-unitary of this form thus modulates the relative contribution of the unitary part of the channel.  Second, varying the time $\tau$ over which the error channel acts changes the relative contributions of the exponential and trigonometric terms in \erf{likelihood}, thus allowing them to be distinguished.  We refer to the first of these control mechanisms as unitary control and the second as $\tau$-variation.

Before proceeding, we note that the likelihood functions~(\ref{likelihood}) are even in $\omega$.  The resulting degeneracy between $\omega$ and $-\omega$ can be lifted by using a counter-unitary $V_c$, but not by changing $\tau$.  Throughout the following, we assume that this ambiguity has been resolved and that $\omega$ is positive.

We now illustrate these concepts with a several examples, all of which assume a prior distribution over $(\omega,\gamma)$ that is a Gaussian with mean $(\omega,\gamma)=(1,0.01)$ and variance $(10^{-2},10^{-6})$, making the ratio of standard deviation to mean equal to $0.1$ for both $\omega$ and $\gamma$ (the special value $\omega=1$ amounts to a choice of temporal units).  The estimation task is to reduce the variances of $\omega$ and $\gamma$.

As we investigate estimation of $\omega$ and $\gamma$ through multiple rounds of error correction, we also explore the preservation of the quantum information in the code subspace.  To make this concrete, we choose a threshold of 5\% for the per-round error probability~(\ref{eq:R}) of uncorrectable errors, which destroy the encoded quantum information.  This threshold, $\Rt=0.05$, corresponds to a threshold on the single-qubit error probability $P\le\Pt=\Ptvalue$.  The dependence of $R$ on $\omega$ and $\tau$, with $\gamma$ fixed at $0.01$, is explored in \frf{fig2}(a).  For the prior mean values, $\omega=1$ and $\gamma=0.01$, we need to perform error correction at times $\tau\le\tau_1=\tauonevalue$ to stay within the threshold.  These values of $\omega$, $\gamma$, and $\tau$ calibrate our analysis.  For these values, $\omega\tau$ and $\gamma\tau$ are small enough that the small-error approximation to $P$ of \erf{eq:P} is fairly good; within this approximation, $\omega$ makes a contribution to $P$ of about $0.13$, nearly 10 times as big as the $0.014$ contribution of $\gamma$.  Both parts of \frf{fig2} illustrate the improvement in quantum-information protection obtained by going from the three-qubit repetition code to the five-qubit repetition code.

\begin{figure*}[ht]
\includegraphics[width=\linewidth]{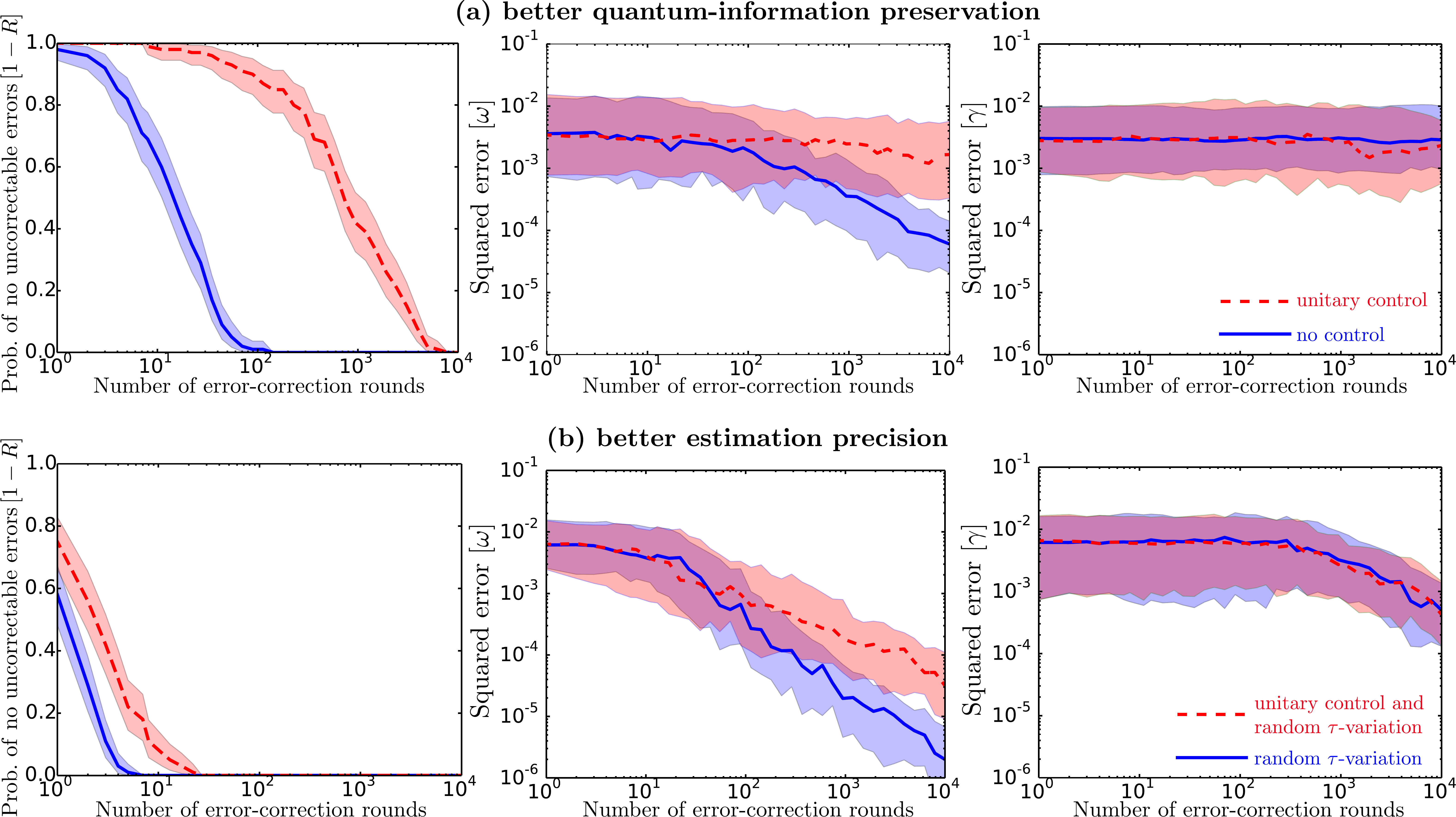}
\caption{\label{fig3}
Tradeoffs between estimation precision and quantum-information preservation.  The plots show results from simulations of $10^4$ rounds of error correction using control strategies described in the text.  For all simulations, the initial distribution of $(\omega,\gamma)$ is normal with mean $(\omega=1,\gamma=0.01)$ and variance $(10^{-2},10^{-6})$.  In the plots the squared error of $\gamma$ is scaled by $10^4$ to put it on the same scale as the squared error of $\omega$.  The plots are assembled from the data accumulated from 100 simulation runs.  A simulation run samples actual values of $\omega$ and $\gamma$ from the prior distribution, which are used to generate syndrome data for successive rounds of error correction.  After each round, Bayesian updating determines a posterior distribution for $\omega$ and $\gamma$; the squared error is the square of the difference between the Bayesian mean of $\omega$ or $\gamma$ given by the posterior and the actual value used for that simulation run.   In the plots of squared error ($\omega$ in the middle and $\gamma$ on the right), the solid or dashed line is the median of the 100 simulation runs, and the surrounding shaded area is the interquartile range (middle 50\% of the 100 simulation runs).  The simulation data are also used to estimate the probability of no uncorrectable errors as a function of round number; in these plots (on the left) the dark line is the mean estimate of this probability, and the shaded area is the 95\% confidence interval.  (a)~The solid blue line for ``no control'' comes from simulations in which $\tau$ is held fixed at $\bar\tau=\taubarvalue$ and there is no counter-unitary ($\omega_c=0$); the red dashed line labeled ``unitary control'' keeps $\tau$ fixed at $\bar\tau$, but uses a  counter-unitary chosen nearly to cancel the $\omega$ rotation, as described in the text.  (b)~The solid blue line labeled ``random $\tau$-variation'' does not use a counter-unitary, but chooses $\tau$ randomly from the interval $[0,1/\gamma]=[0,100]$; the dashed red line labeled ``unitary control and random $\tau$-variation'' adds the counter-unitary strategy described in~(a).}
\end{figure*}

\begin{figure}[ht!]
\includegraphics[width=0.95\linewidth]{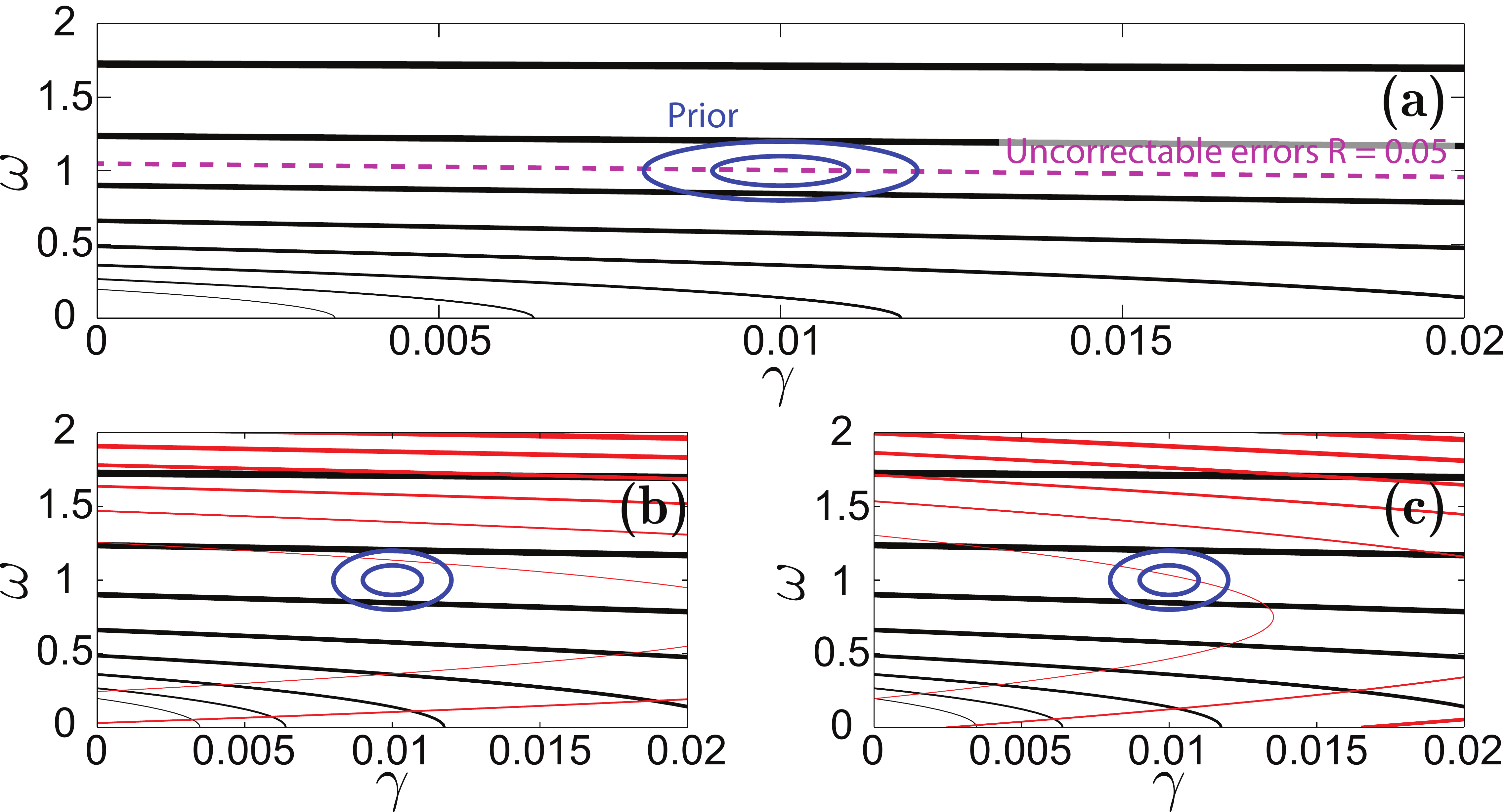}
\caption{\label{fig4}
Graphical representation of what can be learned about $\omega$ and $\gamma$, with and without breaking the degeneracy in the probability $P$ for a bit flip on a single qubit [see \erf{eq:P}]; the likelihoods~(\ref{likelihood}) on which estimation is based are functions only of $P$.  (a)~Solid black lines are contours of the single-qubit error probability $P$ for channel duration {$\tau=\bar\tau=\taubarvalue$}; increasing thickness of the contour lines indicates increasingly larger values of $P$. The dashed purple contour corresponds to our threshold for uncorrectable errors, $\Rt=0.05$.  A sequence of identical rounds of error correction can determine $P$, but cannot distinguish degenerate values of $\omega$ and $\gamma$, i.e., those that give the same value of $P$.  This does not mean that one learns nothing about either $\omega$ or $\gamma$.  Over most of the rectangle of the plot, the contours are nearly horizontal, meaning that $\omega$ makes the dominant contribution to $P$.  As a result, for a flat prior over the entire rectangle of the plot, determining the contour allows one to estimate $\omega$ fairly well, but scarcely provides any information about $\gamma$.  The same can be said for the Gaussian prior used for our simulations, whose contours at one and two standard deviations from the mean are plotted as solid blue lines.  (b)~Contours of $P$ for two values of the counter-rotating control unitary $V_c$: solid black lines are for no control unitary, $\omega_c=0$, as in~(a), and solid red lines for { $\omega_c=0.75$}.  The control unitary displaces the contours along the $\omega$ axis so that the two sets of contours intersect in unique points; a protocol that in many rounds uses both values lifts the degeneracy and determines both $\omega$ and $\gamma$.  (c)~Contours of $P$ for two values of the control parameters $\omega_c$ and $\tau$: solid black lines are as in~(a), i.e., $\omega_c=0$ and $\tau=\tau_1$, and solid red contours are for {$\omega_c=0.75$ and $\tau=2\tau_1$}.  Using these two sets of contours lifts the degeneracy and allows one to determine both $\omega$ and $\tau$.  We note that in both~(b) and~(c), the way the red contours strike across the prior Gaussian means that if one used only the red contours for estimation, one would sacrifice learning about $\omega$ relative to the black contours without picking up much sensitivity to $\gamma$.}
\end{figure}

There are many approaches to combining estimation with error correction. For example, if the channel is static over a reasonable timescale, one could imagine first estimating the channel without trying to preserve any of the quantum information; then, after the channel was estimated, the quantum information could be encoded, and a control counter-unitary applied as in \frf{fig1}(a).  The examples in \frf{fig3} are chosen to illustrate the tradeoffs between learning about the error channel and preserving the quantum information in the code subspace.  For these examples, we use the average probability of a bit flip, $\bar P$, and the average probability of uncorrectable errors, $\bar R$, these being averages of $P$ and $R$ over the Gaussian prior distribution of $\omega$ and $\gamma$.  A fiducial round duration $\bar\tau$ is selected by requiring that $\bar R$ satisfy our threshold condition: $\bar R\le\Rt=0.05$ gives $\tau\le\bar\tau=\taubarvalue$, and this $\bar\tau$ leads to $\bar P=\Pbart=\Pbartvalue$.  Because the prior is quite narrow, $\bar\tau$ and $\Pbart$ are scarecely different from our calibrating values, $\tau_1$ and $\Pt$.  All the examples in \frf{fig3} are carried through $10^4$ rounds of error correction. The stochastic simulations in \frf{fig3} and elsewhere in the manuscript were performed using the Python-based package Qinfer~\cite{QInfer}.

Figure~\ref{fig3}(a) compares two strategies.  For the strategy labeled ``no control,'' we apply no counter-unitary and leave the duration of the error-correction rounds fixed at $\tau=\bar\tau=\taubarvalue$.  With this strategy, we learn next to nothing about $\gamma$, but after enough rounds to make the syndrome data sensitive to $\omega$ at the level of its prior uncertainty, we learn steadily about $\omega$ in subsequent rounds.  The reason for the asymmetry between $\omega$ and $\gamma$ is that given the prior distribution, $\omega$ dominates the errors; we discuss this asymmetry further below and in \frf{fig4}.  Under this strategy, the quantum information is badly degraded after $\sim1/\bar R=1/\Rt=10$ rounds.

The other strategy in \frf{fig3}(a), labeled unitary control, keeps $\tau$ fixed, but uses a counter-unitary in every round, with $\omega_c$ chosen to be minus the current Bayesian mean estimate of $\omega$ plus the current standard deviation of $\omega$~\cite{WieGraFer13}.  The purpose of adding the standard deviation is to avoid nulling out the contribution of $\omega$ to the syndrome measurements; as the estimate of $\omega$ becomes more refined, the standard deviation decreases, and the control should completely correct away the unitary~$U$.  What actually happens in this case, however, is that very little is learned about $\omega$, because the counter-unitary reduces the typical contribution $\omega$ makes to $P$ by a factor of roughly 100, leaving the $\gamma$ error as the dominant contributor to $P$ by a factor of 10.  The consequence is that this strategy gleans little information about either $\omega$ or $\gamma$.  Since the single-qubit error probability is reduced by a factor of 10 relative to the previous strategy, however, the typical $R$ is reduced by a factor of 100, and this means that the encoded quantum information survives for roughly 100 times as many rounds as in the previous strategy.

Figure~\ref{fig3}(b) adds the additional element of changes in the duration $\tau$ of the error-correction rounds.  Our main interest here is to use this $\tau$-variation to learn more rapidly about $\omega$ and $\gamma$.  We could attempt to perform locally optimal adaptive estimation by choosing $\tau$ and $\omega_c$ such that the conditional expected variance of the posterior is maximally decreased, as was done in Ref.~\cite{SerChaCom11}. In this case it is known that the estimation error can be reduced exponentially until the characteristic coherence time is reached~\cite{FerGraCor13}.  Instead, simply to illustrate learning, we choose $\tau$ randomly in each round from the interval $[0,1/\gamma]=[0,100]$, durations that are generally wildly outside the regime of small error probability, making it clear that the quantum information will be degraded very quickly.  For the strategy labeled ``random $\tau$-variation,'' we do not employ any counter-unitaries, and for the strategy labeled ``unitary control and random $\tau$-variation,'' we add counter-unitaries using the same strategy as in \frf{fig3}(a).  For both strategies the syndrome data provides useful information about $\omega$ and $\gamma$.  By reducing the contribution of $\omega$ to $P$, unitary control makes it harder to learn about $\omega$, without much affecting what is learned about $\gamma$, and extends the life of the encoded quantum information.

Figure~\ref{fig4} provides a graphic representation of what can be learned about $\omega$ and $\gamma$ with and without breaking the degeneracy in $P$, providing some intuition for understanding the results of the numerical simulations.

Our protocol's performance, displayed in \frf{fig3}, in protecting quantum information encoded in the code subspace is partly due to the small size of the three-bit repetition code.  The plots in \frf{fig2} indicate that a higher distance code can, for a fixed error-channel duration $\tau$, significantly reduce the probability of a catastrophic error and, by the same token, if $\tau$ is allowed to vary, mean less frequent correction.  This is generically true of higher distance codes and concatenated codes~\cite{RahDohMab02}.

We can explore these advantages analytically by considering the $M$-bit repetition code, which can correct bit-flip errors on up to $m=M'/2-1$ bits, where $M'=M+s$, with $s=0$ if $M$ is even and $s=1$ if $M$ is odd.  Thus the probability per round of an uncorrectable error is
\begin{align}
R&=\prob\bigl[m\ge M'/2\bigr]\nonumber\\
&=\sum_{m=M'/2}^M
\frac{M!}{m!(M-m)!}P^m(1-P)^{M-m}
\end{align}
The Chernoff bound~\cite{chernoff1952,HoeffdingChernoff1963} for a binomial distribution can be used to upper bound $R$. Taking $x=2PM/M'=2P/(1+s/M)$ we find
\begin{align}
R&\le\exp\!\left(-\frac{M}{2}\left(1+{s\over M}\right){(1-x)^2\over1+x}\right)\nonumber\\
&\le\exp\!\left(-\frac{M}{2}\frac{(1-x)^2}{1+x}\right),
\end{align}
The function $(1-x)^2/(1+x)$ decreases monotonically from 1 at $x=0$ to 0 at $x=1$, so if
$x\le\half$, i.e., $P\le\frac14(1+s/M)$, we have $(1-x)^2/(1+x)\ge\frac16$, which gives $R\le e^{-M/12}$.  The probability of no uncorrectable errors after $N$ rounds is $(1-R)^N\ge1-NR\ge1-Ne^{-M/12}$, so the probability $R_N=1-(1-R)^N$ of an uncorrectable error after $N$ rounds satisfies $R_N\le Ne^{-M/12}$ for $P\le\frac14(1+s/M)$.  Thus, for example, if the number of rounds is exponentially large in $M$, but satisfying $N\le e^{M/24}$, the probability of an uncorrectable error, $R_N\le e^{-M/24}$, is exponentially small in~$M$.

There are a number of straightforward extensions of our methods. Perhaps the most important extension is to genuine quantum error-correcting codes, which protect against all single-qubit errors, not just bit flips. We consider such codes in Sec.~\ref{sec:therealthing}, and for now mention other straightforward extensions of our ideas.  An obvious extension is to the case where each qubit in a repetition code experiences a different error channel, characterized by its own unitary strength $\omega$ and decoherence rate $\gamma$, and we have the ability to apply different counter-unitaries to each of the qubits.  Another interesting extension is to allow $\omega(t)$ to vary slowly and deterministically or stochastically in time, so that the control problem becomes tracking $\omega$ and countering its effects as it changes.  From na\"{\i}ve Nyquist-rate arguments, we would need the repetition rate of the error-correction rounds to be at least twice the highest frequency in the bandwith of $\omega(t)$.

\section{Hypothesis testing}\label{sec:Hypothesis_test}
There are numerous hypotheses one might test using error-syndrome data.  In this section, we focus on an example of using syndrome data from the three-qubit repetition code to distinguish an error channel with uncorrelated bit-flip errors from one with correlated bit-flip errors.

To get started, consider the channel $\sF\equiv \sE^{\otimes 3}$, where $\sE$ is the single-qubit bit-flip channel of \erf{eq:Epx}, and the spatially correlated channel
\begin{equation}\label{eq:S}
\sS\equiv \sE_{1,2}\circ \sE_{2,3}\circ\sF\;,
\end{equation}
where
\begin{equation}
\sE_{i,j}(\rho)= (1-q)\rho + q X_i X_j\rho X_j X_i
\end{equation}
describes a simultaneous bit flip of qubits $i$ and $j$ with probability $q$.  The indices $i,j$ on the $X$ Pauli operators indicate which qubit they act on.  The channels $\sE_{1,2}$ and $\sE_{2,3}$ introduce correlated errors; when $q = 0$, $\sS$ reduces to $\sF$.  Quantum circuits for these two channels are depicted in \frf{fig6}.

The syndrome probabilities for the uncorrelated-error channel $\sF$ are
\begin{subequations}\label{null_hyp}
\begin{align}
\prob(00|p) &= 1-3p(1-p),\\
\prob(S|p)  &= p(1-p),
\end{align}
\end{subequations}
where $S\in \{01,10 ,11\}$ [same as \erf{likelihood} with $\omega=0$].  We record in \arf{app:corr} the error probabilities for $\sS$; these lead to the syndrome probabilities of \erf{eq:H1syndromeprobs}, which we repeat here in the form
\begin{subequations}\label{alt_hyp}
\begin{align}
\prob(00|p,q) &= 1-3p(1-p)-q(2-q)(1-2p)^2,\\
\prob(10|p,q) &=  p(1-p)+q(1-q)(1-2 p)^2,\\
\prob(01|p,q) &=  p(1-p)+q(1-q)(1-2 p)^2,\\
\prob(11|p,q) &=  p(1-p)+q^2(1-2 p)^2.
\end{align}
\end{subequations}
Setting $q=0$ in these expressions reduces them to the syndrome probabilities~(\ref{null_hyp}) for $\sF$.

The two hypotheses or models we consider are the following.  The uncorrelated-error hypothesis, which we label $\Hzero$, is that the error channel is $\sF$ with an unknown error probability $p$ drawn from a distribution $\prob(p)$.  The correlated-error hypothesis, labeled $\Hone$, is that the error channel is $\sS$ with unknown probabilities $p$ and $q$, $p$ being drawn from $\prob(p)$ and $q$ from a distribution $\prob(q)$.  In the simulations presented in \frf{fig7}, both $\prob(p)$ and $\prob(q)$ are flat on the interval $[0,0.1]$.

To put the notation for the two hypotheses on the same footing, we say that the model $\Hzero$ draws $p$ and $q$ from a distribution $\prob(p,q|\Hzero)=\prob(p)\delta(q)$, and the model $\Hone$ draws $p$ and $q$ from a distribution $\prob(p,q|\Hone)=\prob(p)\prob(q)$.  The syndrome probabilities for the two hypotheses, which now become likelihood functions, are
\begin{subequations}\label{eq:syndromeHzeroHone}
\begin{align}\label{eq:syndromeHzero}
\prob(S|\Hzero)
&=\int dp\,dq\,\prob(S|p)\prob(p,q|\Hzero)\nonumber\\
&=\int dp\,\prob(S|p)\prob(p),\\
\prob(S|\Hone)&=\int dp\,dq\,\prob(S|p,q)\prob(p,q|\Hone)\nonumber\\
&=\int dp\,dq\,\prob(S|p,q)\prob(p)\prob(q),
\label{eq:syndromeHone}
\end{align}
\end{subequations}
where $S$ here stands for any of the four syndromes.

\begin{figure}[ht]
\includegraphics[width=0.9\linewidth]{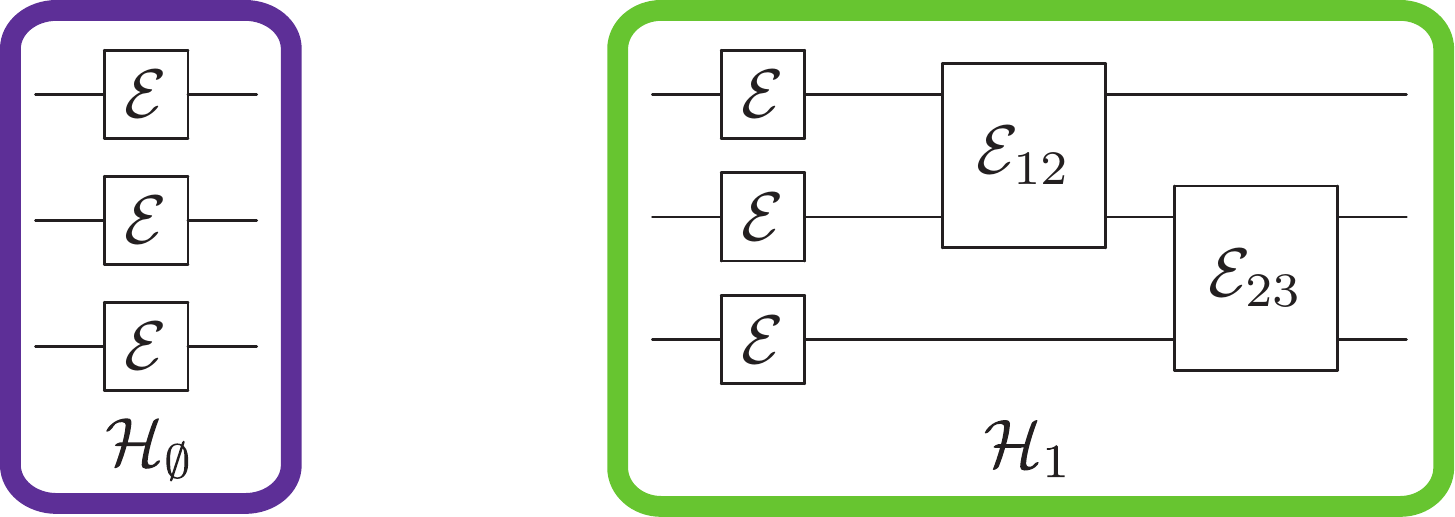}
\caption{\label{fig6}
Quantum circuits for the alternatives in our hypothesis testing of correlated vs.~uncorrelated noise. (a)~Uncorrelated-error channel $\sF=\sE^{\otimes 3}$, which is parameterized by the single-qubit bit-flip probability $p$.  The null hypothesis $\Hzero$ is the channel $\sF$ with $p$ drawn from the distribution $\prob(p)$.  The syndrome probabilities are given by Eqs.~(\ref{null_hyp}) and~(\ref{eq:syndromeHzero}).  (b)~Spatially correlated error channel $\sS=\sE_{1,2}\circ\sE_{2,3}\circ\sF$; $\sE_{i,j}$ describes a simultaneous bit flip of qubits $i$ and $j$ with probability~$q$.  The channel $\sS$ is characterized by parameters $p$ and $q$ and reduces to $\sF$ when $q=0$.  The alternative hypothesis $\Hone$ is the channel $\sS$ with $p$ drawn from $\prob(p)$ and $q$ from a distribution $\prob(q)$.  The syndrome probabilities are given by Eqs.~(\ref{alt_hyp}) and~(\ref{eq:syndromeHone}).}
\end{figure}

To perform hypothesis testing, we use \emph{model selection\/} and compare the Bayesian posterior probabilities assigned to $\Hzero$ and $\Hone$ given the data $\data$ collected from rounds of syndrome measurements.  The Bayesian posteriors are
\begin{subequations}
\begin{align}
\prob(\Hzero|\data)&=\frac{\prob(\data|\Hzero)\prob(\Hzero)}{\prob(\data)}\;,\\
\prob(\Hone|\data)&=\frac{\prob(\data|\Hone)\prob(\Hone)}{\prob(\data)},
\end{align}
\end{subequations}
where $\prob(\Hzero)$ and $\prob(\Hone)$ are prior probabilities and $\prob(D)$ is the unconditioned probability of the data.  For equal prior probabilities, which we assume henceforth to avoid any prejudice for either of the alternatives, the posteriors reduce to
\begin{subequations}\label{eq:posteriors}
\begin{align}
\prob(\Hzero|\data)
&=\frac{\prob(\data|\Hzero)}{\prob(\data|\Hzero)+\prob(\data|\Hone)},\\
\prob(\Hone|\data)
&=\frac{\prob(\data|\Hone)}{\prob(\data|\Hzero)+\prob(\data|\Hone)}.
\end{align}
\end{subequations}

What we want to do is to assess the syndrome data's efficacy at confirming the ``true'' hypothesis.  For that purpose, imagine that the data are generated by the hypothesis $\Hzero$.  In this circumstance, we want measures of how good the posterior probabilities are at confirming $\Hzero$.  An appropriate measures is the median value of the posterior probabilities $\prob(\Hzero|D)$ and $\prob(\Hone|D)$ when the data is generated by $\Hzero$.  These median values can be thought of as probabilities to confirm $\Hzero$ or $\Hone$ given that $\Hzero$ is true.  The same considerations apply when $\Hone$ is considered to be the true hypothesis.

\begin{figure}[ht]
\includegraphics[width=\linewidth]{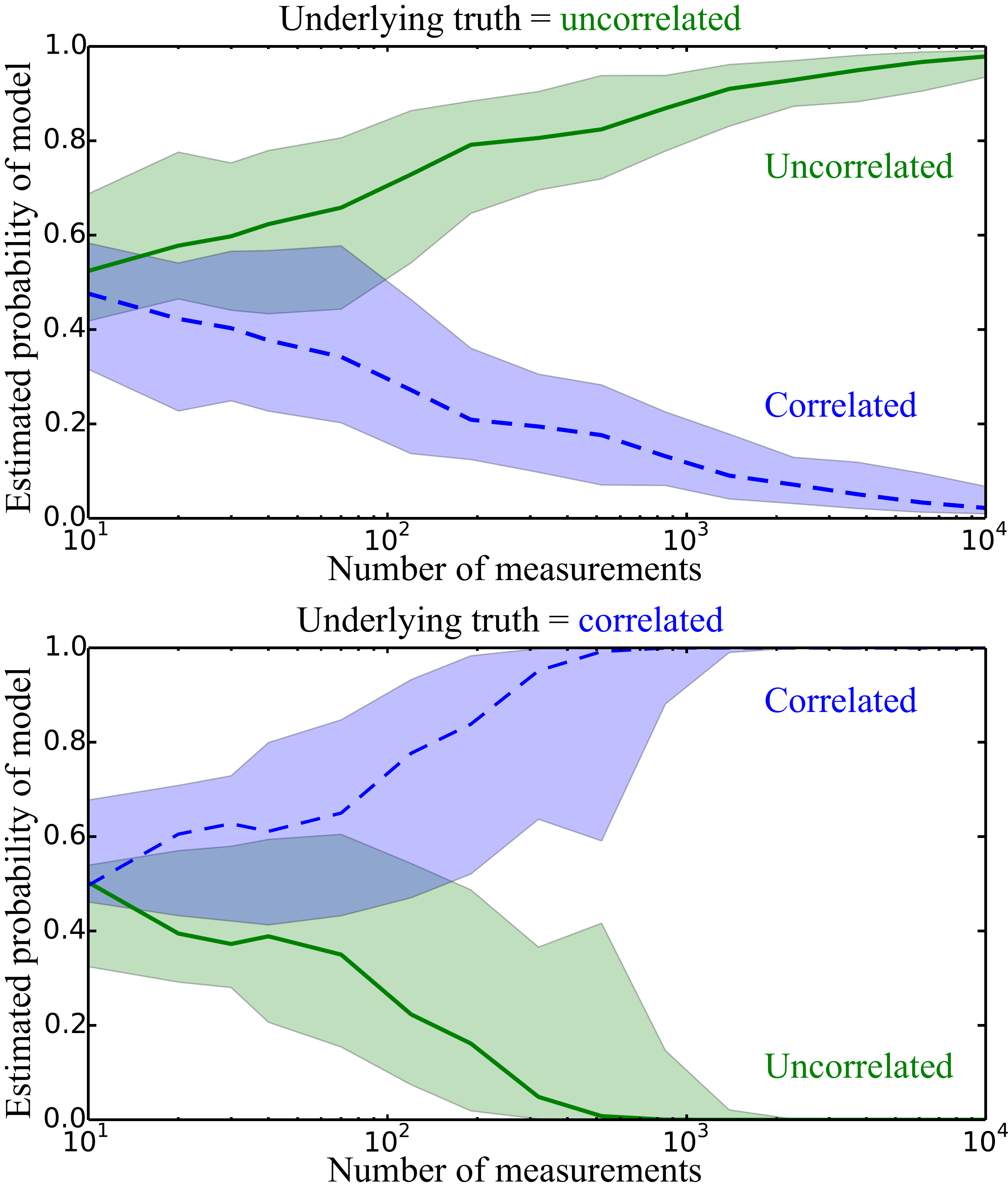}
\caption{\label{fig7}
Simulations of Bayesian posterior probabilities for the uncorrelated-error hypothesis $\Hzero$ (green), which is defined by the syndrome probabilities~(\ref{null_hyp}) with $p$ drawn randomly from the interval $[0,0.1]$, and the correlated-error hypothesis $\Hone$ (blue), which is defined by the syndrome probabilities~(\ref{alt_hyp}) with $p$ and $q$ both drawn randomly from the interval $[0,0.1]$.   In the simulation runs, the data are generated by a ``true'' hypothesis.  When $\Hzero$ is true (top plot), a simulation run consists of drawing $p$ randomly from $[0,0.1]$, generating syndrome data according to Eqs.~(\ref{null_hyp}), and calculating the Bayesian posteriors~(\ref{eq:posteriors}).   When $\Hone$ is true (bottom plot), a simulation run consists of drawing both $p$ and $q$ randomly from $[0,0.1]$, generating syndrome data according to Eqs.~(\ref{alt_hyp}), and calculating the Bayesian posteriors~(\ref{eq:posteriors}).  Both plots are based on 100 simulation runs.  The solid line is the median of the posterior probability over the runs, and the shaded region is the interquartile range of the 100 runs.  The data in the top (bottom) plot provide estimates of the posterior probabilities~\ref{eq:posteriors} when $\Hzero$ ($\Hone$) is true.}
\end{figure}

Figure~\ref{fig7} summarizes the results of simulations that estimate the posterior probabilities in the situation where $\prob(p)$ and $\prob(q)$ are flat on the interval $[0,0.1]$.  These results confirm that, for hypotheses defined by this range of values of $p$ and $q$, the model-selection approach converges to the ``true'' hypothesis after a few hundred rounds, regardless of which hypothesis is regarded as true.

This is not surprising.  For small, roughly equal values of $p$ and $q$, the dominant contributions to the syndrome probabilities~(\ref{alt_hyp}) are $\prob(10|p,q)=\prob(01|p,q)=p+q$ and $\prob(11|p,q)=p$; $N$ rounds of error correction allow one to estimate these syndrome probabilities with an accuracy of about $\sqrt{p/N}$, so the two hypotheses become distinguishable when $\sqrt{p/N}\sim q$, i.e., $N\sim p/q^2\sim 1/p$.  For $p=0.05$, in the middle of its range, this yields $N\sim200$, consistent with the full Bayesian simulations of \frf{fig7}.

The general idea of this section is that error models can be discriminated using syndrome data if the models have different syndrome statistics.  Just as in the considerations of parameter estimation in \srf{sec:Param_est}, one might be able to use control mechanisms to make the syndrome probabilities more or less distinguishable.  For example, in the situation considered in this section, if one modeled how $p$ and $q$ depend on the error-correction time $\tau$, one could vary $\tau$ to make it easier to discriminate the two models.

We close this section by pointing interested readers to the work of Schwarz and van Enk~\cite{SchvanEnk13}, who consider a form of model selection based on the Akaike information criterion.  Schwarz and van Enk use this form of model selection to determine when a parsimonious error model is not sufficient to characterize the statistics one is \hbox{observing}.

\section{Generalizing to genuine quantum codes}\label{sec:therealthing}

Our analysis until now has been restricted to a setting where the only errors are bit flips.  This allows us to use the simple repetition code to explore how our techniques work.  In a realistic setting, however, we must allow for a wider class of errors, and thus we should use more general, genuinely quantum-mechanical codes that correct a wider class of errors.  In this section we sketch how the ideas of \srf{sec:Param_est} can be applied to more realistic channels and to more general codes.  As an example of the latter, we give syndrome probabilities for the five-qubit ``perfect'' quantum code that corrects all single-qubit errors~\cite{LafMiqPazZur96,BenDiVSmoWoo96}.

The error channel we consider has independent errors on all the qubits of an $M$-qubit code; as in \srf{sec:Param_est}, the error channel has a nonunitary piece and a unitary piece.  The nonunitary map for each qubit is the anisotropic error channel,
\begin{align}\label{eq:Exyz}
\sE(\rho) =& (1-q)I \rho I + p_x X\rho X+p_y Y\rho  Y+p_z Z\rho Z,
\end{align}
where $p_x$, $p_y$, and $p_z$ are the probabilities for a bit flip ($X$), a bit-phase flip ($Y$), and a phase flip ($Z$), and where $q=p_x+p_y+p_z$.  The unitary party of the channel is the map $\sU(\rho)=U \rho U^\dagger$, where
\begin{align}\label{eq:Ugenrot}
U=e^{-i\boldsymbol{n}\cdot\boldsymbol{\sigma}\theta/2}
=I\cos(\theta/2)-i\boldsymbol{n}\cdot\boldsymbol{\sigma}\sin(\theta/2)
\end{align}
is a rotation by angle $\theta=\omega\tau$ about axis $\boldsymbol{n}$.  The total channel for each qubit is $\sU\circ\sE$, and the channel for the $M$ physical qubits is $\sF=\sU^{\otimes M}\circ\sE^{\otimes M}$.

We give the explicit form of the channel $\sU\circ\sE(\rho)$ in \arf{app:anisotropic}.  The explicit form has many terms, but many of these can be ignored for any code whose syndrome measurements force the qubits to commit to Pauli errors, as do all stabilizer codes.  The only terms that we need to retain are those that correspond to no error (terms proportional to $\rho$), to bit flips ($X\rho X$),  bit-phase flips ($Y\rho Y$), and phase flips ($Z\rho Z$).  The constants multiplying these terms are the probability $Q$ for no error and the probabilities $P_x$, $P_y$, and $P_z$ for $X$, $Y$, and $Z$ errors.  Discarding the irrelevant terms, we are left with an effective channel
\begin{align}\label{eq:Pxyzchannel}
\sG(\rho)=
&Q\,\rho + P_x X\rho X +P_y Y\rho Y +  P_z Z\rho Z,
\end{align}
where
\begin{subequations}\label{eq:QPxyz}
\begin{align}
Q   &= (1-q)c^2 + (p_x n_x^2+p_y n_y^2+p_z n_z^2)s^2,\\
P_x &= p_x c^2 + [(1-q) n_x^2+p_y n_z^2+p_z n_y^2]s^2,\\
P_y &= p_y c^2 + [(1-q) n_y^2+p_x n_z^2+p_z n_x^2]s^2,\\
P_z &= p_z c^2 + [(1-q) n_z^2+p_x n_y^2+p_y n_x^2]s^2.
\end{align}
\end{subequations}
Here we use the abbreviations $c=\cos(\theta/2)$ and $s=\sin(\theta/2)$.

Before proceeding, notice that $\sU$ and $\sE$ do not commute, so their order in $\sF$ is important.  We could consider having the unitary and nonunitary parts of the channel act simultaneously and describe that situation by a generalization of the vacuum master equation~(\ref{eq:vacuum_master}) in which the Hamiltonian rotates about $\boldsymbol n$ and there are separate rates for the three Pauli flips.  As we show in \arf{app:anisotropic}, the resulting channel acting for a duration $\tau$ can be written in terms of a time-ordered solution for the Bloch vector, and from this Bloch-vector solution, an effective channel of the form~(\ref{eq:Pxyzchannel}) can be extracted.

The probabilities~(\ref{eq:QPxyz}) determine everything about strings of errors on the $M$ qubits.  Since the qubits are subject to independent errors, if we want to know the probability of a string that specifies for each qubit either no error or a particular flip, we simply multiply together the appropriate probabilities from Eqs.~(\ref{eq:QPxyz}).  In this way, for example, the probability for $j_x$ $X$ errors, $j_y$ $Y$ errors, and $j_z$ $Z$ errors is given by a standard multinomial distribution,
\begin{align}
P_{j_x,j_y,j_z}=\frac{M!}{j_x!j_y!j_z!J!}P_x^{j_x}P_y^{j_y}P_z^{j_z}Q^J,
\end{align}
where $Q=1-P_x-P_y-P_z$ and $J=M-j_x-j_y-j_z$.

How the probabilities for error strings enter into the syndrome probabilities will vary from code to code, but the syndrome probabilities will always be combinations of the probabilities $P_x$, $P_y$, and $P_z$.   For the perfect five-qubit code~\cite{LafMiqPazZur96,BenDiVSmoWoo96}, which can correct all single-qubit errors, the 16 syndromes, corresponding to the results of four parity measurements, fall into four classes, $S_0=\{0000\}$, $S_1=\{0001,0011,0110,1000,1100\}$, $S_2=\{0010,0100,0101,1001,1010\}$, and $S_3=\{0111,1011,1101,1110,1111\}$.  The syndrome probabilities are constant on each class and are functions of $P_x$, $P_y$, and $P_z$; we give the explicit form of the syndrome probabilities for the five-qubit code in \arf{app:messy}.

In the absence of control mechanisms, the syndrome data from many rounds of error correction allow us to determine only the three probabilities $P_x$, $P_y$, and $P_z$, and this falls short of determining all the properties of the channel, specifically, the strength $\omega$ and axis $\boldsymbol{n}$ of the systematic rotation and the error probabilities $p_x$, $p_y$, and $p_z$.  The problem, as in the simple analysis of the three-qubit repetition code in \srf{sec:Param_est}, is degeneracies, in this case, degeneracies within $P_x$, $P_y$, and $P_z$.  We do, however, have available many control mechanisms, particularly the ability to perform counter-unitaries of arbitrary strength and arbitrary axis on each of the five qubits and to vary the time over which the channel acts.  These, if used judiciously, will allow us, as in the simple analysis, to lift the degeneracies and to determine all the parameters of the channel.

\section{Relationship to prior work}\label{sec:prior}

In this section we briefly consider previous, related work on the use of error-correcting codes for metrological and estimation tasks.

The methods we describe here might find application to quantum sensors, which are designed to determine a classical parameter as precisely as possible given the quantum-mechanical properties of the sensor. In the metrology of quantum sensors, one typically seeks to estimate classical parameters that are impressed by some quantum dynamics on a series of quantum systems prepared in the same initial state.  Estimation of the parameters $\omega$ and $\gamma$ of an error channel, as in \srf{sec:Param_est}, is an example of such quantum metrology.

In contrast to traditional quantum metrology, which envisions probing a sequence of identically prepared systems, our estimation techniques can be done on a single encoded qubit, which is automatically re-prepared in the logical subspace at the end of each error-correction round. Information about the desired parameter is extracted from syndrome data---and nothing else---by using the control mechanisms we describe here to distinguish unitary errors from random errors; error correction automatically re-prepares the system in an appropriate initial state after each round, ready for the next round.  Also, and again unlike conventional metrology, any state in the code space, be it mixed or pure, has the same estimation efficacy; i.e., we have inherent robustness to some noise on the probe state.

Recently, a related series of papers~\cite{DurSkoFroKra13,Ozeri13,KesLovSusLuk13,ArrVinAhaRet13} proposed using error-correcting codes to preserve Heisenberg scaling in quantum metrology using lossy, decoherent systems.  This possibility was pointed out by Preskill~\cite{Preskill} over a decade ago.  The essential technique used by these approaches is to impart the parameter to be estimated via a logical rotation, which rotates the encoded state in the code subspace.  The errors take the system out of the logical subspace and thus can be corrected without affecting the signal.  The signal is ultimately detected by probing the state change produced by the parameter in the code subspace.  Three of these papers~\cite{DurSkoFroKra13,Ozeri13,ArrVinAhaRet13} employ quantum codes that can correct any single-qubit error, which requires that the signal be imparted by a high-weight logical operator.  The fourth~\cite{KesLovSusLuk13} uses a code that only corrects bit-flip errors and imparts the signal via single-qubit logical operators.

We do something quite different in this paper: both the unitary signal and the random errors are errors; both take the system out of the code subspace.  The unitary errors are distinguished from the random errors by using control mechanisms to make the unitary and random-error contributions to the syndrome statistics different; learning about either type of error requires only the syndrome measurements that are performed in the course of quantum error correction.

Two very recent papers posted to the arXiv e-print server~\cite{FowSankKelBarMar2014,OmSriBan2014} study using syndrome-measurement data to perform estimation tasks.  In contrast to these two papers, in the work reported here, we focus on procedures that can be performed using syndrome data, including model selection and parameter estimation that use control mechanisms to separate unitary and random errors; in addition, we perform numerical simulations to demonstrate and characterize the performance of these procedures, including the tradeoff between accuracy and the protection of encoded quantum information.

In Ref.~\cite{FowSankKelBarMar2014}, the authors focus on scalable characterization of the ways that errors compose and propagate in specific circuits for quantum error detection; the emphasis is on estimating important parameters of Pauli error channels from syndrome data, without the attention to distinguishing different contributions to error channels that occupies us in {\srf{sec:Param_est}}.

Reference~\cite{OmSriBan2014} studies the problem of performing quantum process tomography in an ``online'' setting, demonstrating how to determine from syndrome data the elements of the process matrix that correspond to high-probability, correctable Pauli errors.  To get at the off-diagonal elements of the process matrix, which can be produced by unitary errors, the authors use control unitaries applied between the action of the error channel and the extraction of syndrome data.  This kind of control, which we do not consider in this paper, is another way of lifting the degeneracies between unitary and random errors; it has the disadvantage of destroying the encoded quantum information even without uncorrectable errors.

\section{Discussion}\label{sec:discussion}

We have demonstrated several ways in which the data gathered from error-correction syndrome measurements can be processed by an agent and used intelligently to estimate and discriminate between error channels and to control parts of an error channel.  We close with a brief discussion of extensions and generalizations of the ideas presented in this paper as well as areas requiring future work.

One possibility for the online learning techniques described above is the adaptive modification of the error-correcting code being used.  For example, the code distance of topological codes, a particular family of quantum-error-correcting codes, is related to the geometric dimensions of a lattice of qubits.  This distance can be increased on the fly using only local unitary operations and a reservoir of fresh ancilla qubits.  If, in the process of performing error correction and learning about the noise channel, one learns that a bit-flip error is more likely than a phase-flip error, the code could be adjusted by increasing the size of a particular lattice dimension.

Another application might be the use of an adaptive decoding procedure.  Algorithms for decoding the syndrome data for a quantum code, a process that determines the proper corrective action to take given the syndrome, could take advantage of knowledge that a particular qubit or region of qubits have a higher noise rate.  Incorporating this information into the decoding procedure might result in a different inferred correction and has the potential to increase the code's error threshold.

Perhaps the most important area for future work is to investigate the extent to which our in-situ techniques for device characterization and calibration can be performed in a fault-tolerant setting with experimentally relevant noise models.  Moreover, it will be important to determine the minimal additional resources required for full characterization.

We finish with an shortcoming of our technique.  Our method, like any method which attempts to characterize quantum processes (e.g. process tomography~\cite{MerGamSmo13}), is marred by the curse of dimensionality.  We hope that the methods developed in other parts of the community, such as direct fidelity estimation~\cite{FlaLui11,daSlLanPou11} and tomographic randomized benchmarking \cite{KimdaSRya14}, can be fused with the techniques we present here.


\acknowledgments
The authors gratefully acknowledge productive discussions with Bryan Eastin, Chris Granade, and Matthias Lang. JC, CF, CC, and CMC were supported in part by National Science Foundation Grant No.~PHY-1212445 and by \hbox{Office} of Naval Research Grant No.~N00014-11-1-0082.  CF was supported by the Canadian Government through the NSERC PDF program. MT and HJB acknowledge support from the Austrian Science Fund (FWF) through the SFB FoQuS:\,F\,4012.  GJM  acknowledges the support from the Australian Research Council's Centre of Excellence in Engineered Quantum Systems CE110001013 and the Templeton World Charity Fund grant TWCF0064/AB38.  JC, MT, GJM, HJB, and CMC also acknowledge support from the National Science Foundation under Grant No.~NSF PHY-1125915 at the Kavli Institute for Theoretical Physics; the idea for this paper germinated at the KITP and then branched in several directions.

\begin{widetext}
\appendix

\section{Error probabilities for correlated error model}\label{app:corr}

The error probabilities for the spatially correlated error model $\sS$ of \erf{eq:S} are
\begin{subequations}
\label{eq:H1errprobs}
\begin{align}
\hbox{Pr}(000|p,q)&=(1-p)^3(1-q)^2+p^2(1-p)q(2-q),\\
\hbox{Pr}(100|p,q)=\hbox{Pr}(001|p,q)&=p(1-p)^2(1-q+q^2)+p^3q(1-q),\\
\hbox{Pr}(010|p,q)&=p(1-p)^2(1-q^2)+p^3q^2,\\
\hbox{Pr}(110|p,q)=\hbox{Pr}(011|p,q)&=(1-p)^3q(1-q)+p^2(1-p)(1-q+q^2),\\
\hbox{Pr}(101|p,q)&=(1-p)^3q^2+p^2(1-p)(1-q^2),\\
\hbox{Pr}(111|p,q)&=p(1-p)^2q(2-q)+p^3(1-q)^2,
\end{align}
\end{subequations}
where the three slots are for the three qubits, with 0 denoting no error and 1 denoting an error.  The probability for $m$ bit flips on the three qubits, $p^e_m$, is
\begin{subequations}\label{eq:H1mflipprobs}
\begin{align}
p^e_0 &= \hbox{Pr}(000|p,q)=(1-p)^3(1-q)^2+p^2(1-p)q(2-q),\\
p^e_1 &= \hbox{Pr}(100|p,q)+\hbox{Pr}(010|p,q)+\hbox{Pr}(001|p,q)=p(1-p)^2[3-q(2-q)]+p^3q(2-q),\\
p^e_2 &= \hbox{Pr}(110|p,q)+\hbox{Pr}(101)+\hbox{Pr}(011|p,q)=(1-p)^3q(2-q)+p^2(1-p)[3-q(2-q)],\\
p^e_3 &= \hbox{Pr}(111|p,q)=p(1-p)^2q(2-q)+p^3(1-q)^2,
\end{align}
\end{subequations}
The syndrome probabilities are
\begin{subequations}\label{eq:H1syndromeprobs}
\begin{align}
\prob(00|p,q) &= \hbox{Pr}(000|p,q)+\hbox{Pr}(111|p,q)=[1-3p(1-p)](1-q)^2+p(1-p)q(2-q),\\
\prob(10|p,q) &= \hbox{Pr}(100|p,q)+\hbox{Pr}(011|p,q)=[1-3p(1-p)]q(1-q)+p(1-p)(1-q+q^2),\\
\prob(01|p,q) &= \hbox{Pr}(001|p,q)+\hbox{Pr}(110|p,q)=[1-3p(1-p)]q(1-q)+p(1-p)(1-q+q^2),\\
\prob(11|p,q) &= \hbox{Pr}(010|p,q)+\hbox{Pr}(101|p,q)=[1-3p(1-p)]q^2+p(1-p)(1-q^2).
\end{align}
\end{subequations}

\section{Anisotropic error channel}\label{app:anisotropic}

\subsection{Single-qubit error probabilities for channel $\sU\circ\sE$}

The single-qubit channel of \srf{sec:therealthing} is a composition of the anisotropic error channel $\sE$ of \erf{eq:Exyz} and the rotation~(\ref{eq:Ugenrot}) by angle $\theta=\omega\tau$ about $\boldsymbol n$.  The explicit form of the composite channel is
\begin{align}\label{eq:UE}
\sU\circ\sE
&=\Big[(1-q)c^2+s^2\sum_jp_jn_j^2\Big]\rho
+c^2\sum_jp_j\sigma_j\rho\,\sigma_j
+(1-q)s^2\boldsymbol{n}\cdot\boldsymbol{\sigma}\rho\,\boldsymbol{n}\cdot\boldsymbol{\sigma}
+\sum_{j,k,l,m,n}s^2\epsilon_{jlm}\epsilon_{kln}p_ln_mn_n\sigma_j\rho\,\sigma_k\nonumber\\
&\quad
+sc\sum_{j,k,l}\epsilon_{jkl}n_j(p_k-p_l)\sigma_k\rho\,\sigma_l
-i(1-q)sc\,[\boldsymbol{n}\cdot\boldsymbol{\sigma},\rho]
-isc\sum_jp_jn_j[\rho,\sigma_j]
+is^2\sum_{j,k,l}\epsilon_{jkl}p_jn_jn_k[\rho,\sigma_l],
\end{align}
where $c=\cos(\theta/2)$ and $s=\sin(\theta/2)$.  For a quantum code that commits the qubits to Pauli errors, the only terms that matter in this expression are those for no error (proportional to $\rho$) and those for Pauli errors (proportional to $\sigma_j\rho\sigma_j$).  All these terms arise from the four terms on the top line of \erf{eq:UE}; discarding the terms that don't matter, we are left with the effective channel
\begin{equation}\label{eq:G}
\sG(\rho)
=\Big[(1-q)c^2+s^2\sum_jp_jn_j^2\Big]\rho
+\sum_j\sigma_j\rho\sigma_j
\Big[c^2p_j+s^2\Big((1-q)n_j^2+\sum_{l,m,n}\epsilon_{jlm}\epsilon_{jln}p_ln_mn_n\Big)\Big].
\end{equation}
This effective channel is written in an equivalent form in Eqs.~(\ref{eq:Pxyzchannel}) and (\ref{eq:QPxyz}).

\subsection{Choi transformation for qubit channels}

Another way to get at the same error probabilities is to consider any trace-preserving qubit quantum operation~$\sB$~\cite{MikeandIke}.  Such a  quantum operation is specified by how it acts on the unit operator $I=\sigma_0$ and the three Pauli operators $\sigma_j$, $j=1,2,3$.  Letting Greek indices take on values $0,1,2,3$, we have that $\sB$ is specified by the \emph{transformation matrix\/}
\begin{equation}
\sB_{\alpha\beta}=\frac12\tr\big(\sigma_\alpha\sB(\sigma_\beta)\big).
\end{equation}
That $\sB$ is trace-preserving implies that $\sB_{0\alpha}=0$.

If $\sB$ is also unital, i.e., maps the unit operator to itself, then $\sB_{j0}=0$.  Any unitary channel is unital, and the anisotropic error channel $\sE$ of \erf{eq:Exyz} is also unital.  A unital transformation is specified by the $3\times3$ matrix $B$ whose elements are
\begin{equation}
B_{jk}=\sB_{jk}=\frac12\tr\big(\sigma_j\sB(\sigma_k)\big).
\end{equation}
A quantum state $\rho=\half\big(I+\boldsymbol{S}\cdot\boldsymbol{\sigma}\big)$, with Bloch vector $\boldsymbol{S}=\tr(\rho\boldsymbol{\sigma})$, transforms under a unital $\sB$ according to
\begin{equation}
\sB(\rho)=\frac12\big(I+\sB(\boldsymbol\sigma\cdot\boldsymbol{S})\big)
=\frac12\big(I+\boldsymbol\sigma\cdot B\boldsymbol{S}\big);
\end{equation}
i.e., $B$ is the matrix that transforms the Bloch vector.

We want to get $\sB$ into the form
\begin{equation}
\sB=\frac12\sum_{\alpha,\beta}\chi_{\alpha\beta}\sigma_\alpha\odot\sigma_\beta,
\end{equation}
where here the $\odot$ can be regarded as a place-holder for the operator that $\sB$ acts on.  The matrix $\chi_{\alpha\beta}$ is called the \emph{process matrix\/}; it is Hermitian, and its diagonal elements give the probabilities for Pauli errors.

The transformation from $\sB_{\alpha\beta}$ to $\chi_{\alpha\beta}$, called the \emph{Choi transformation\/}~\cite{Choi75}, has a particularly simple form for qubit operations.  To find the transformation, introduce a superoperator $\sB^\#$ that reverses the roles of the transformation and process matrices, i.e.,
\begin{equation}
\sB^\#=\frac12\sum_{\alpha,\beta}\sB_{\alpha\beta}\sigma_\alpha\odot\sigma_\beta.
\end{equation}
Then the process matrix is given by
\begin{equation}\label{eq:Choi}
\chi_{\alpha\beta}
=\frac12\tr\big(\sigma_\alpha\sB^\#(\sigma_\beta)\big)
=\frac14\sum_{\gamma,\delta}\sB_{\alpha\beta}\tr(\sigma_\alpha\sigma_\gamma\sigma_\beta\sigma_\delta).
\end{equation}

For a unital $\sB$, the Choi transformation~(\ref{eq:Choi}) reduces to
\begin{equation}
\chi_{\alpha\beta}=\frac12\delta_{\alpha\beta}+\frac14\sum_{l,m}B_{lm}\tr(\sigma_\alpha\sigma_l\sigma_\beta\sigma_m).
\end{equation}
Working out the trace gives the explicit form
\begin{subequations}\label{eq:Choiunital}
\begin{align}
\chi_{00}&=\frac12\big(1+\tr(B)\big),\\
\chi_{0j}=-\chi_{j0}&=-\frac{i}{2}\sum_{k,l}\epsilon_{jkl}B_{kl},\\
\chi_{jk}&=\frac12\big(\delta_{jk}+B_{jk}+B_{kj}-\tr(B)\big).
\end{align}
\end{subequations}
The diagonal elements give the probabilities for $x$, $y$, and $z$ errors:
\begin{subequations}\label{eq:ChoiQPxyz}
\begin{align}
Q=1-\sum_jP_j&=\frac12\chi_{00}=\frac14\big(1+\tr(B)\big),\\
P_j&=\frac12\chi_{jj}=\frac14\big(1+2B_{jj}-\tr(B)\big).
\end{align}
\end{subequations}

For the unital channel $\sB=\sU\circ\sE$, the Bloch-vector transformation matrix $E$ corresponding to the anisotropic error channel $\sE$ has matrix elements $E_{jk}=(1-2q+2p_j)\delta_{jk}$, and the Bloch-vector transformation matrix $R_{\boldsymbol n}(\theta)$ corresponding to the unitary channel $\sU$ is the three-dimensional rotation matrix for a rotation by angle $\theta$ about axis $\boldsymbol n$, which has matrix elements $R_{jk}=\delta_{jk}\cos\theta+n_jn_k(1-\cos\theta)-\sin\theta\sum_l\epsilon_{jkl}n_l$.  The overall Bloch-vector transformation matrix is $B=R_{\boldsymbol n}(\theta)E$.  Choi-transforming $B$ according to Eqs.~(\ref{eq:Choiunital}) to get the process matrix gives the channel~(\ref{eq:UE}) and, in particular, the probabilities~(\ref{eq:QPxyz}).

\subsection{Simultaneous flip errors and rotation about arbitrary axis}

Consider now the case where the anisotropic errors and the rotation about $\boldsymbol n$ act simultaneously.  The resulting evolution is described by the master equation
\begin{align}
d\rho &= -i\smallfrac{1}{2}\omega\,dt\,[\boldsymbol{n}\cdot\boldsymbol{\sigma},\rho]
+ 2\,dt\,\sum_j\gamma_j(\sigma_j\rho\sigma_j-\rho),
\end{align}
which generalizes the master equation~(\ref{eq:vacuum_master}) to this situation.  The corresponding Bloch vector $\boldsymbol{S}=\tr(\rho\boldsymbol{\sigma})$ evolves according to
\begin{equation}\label{eq:Blochevolution}
\frac{d\boldsymbol{S}}{dt}
=\omega\boldsymbol{n}\times\boldsymbol{S}
-4\Big(\gamma\boldsymbol{S}-\sum_j\gamma_jS_j\boldsymbol{e}_j\Big)
=\omega\boldsymbol{n}\times\boldsymbol{S}
-4\Gamma\boldsymbol{S},
\end{equation}
where $\gamma=\gamma_x+\gamma_y+\gamma_z$ is the sum of the three error rates and
\begin{equation}
\Gamma=
\begin{pmatrix}
\gamma_y+\gamma_z&0&0\\
0&\gamma_x+\gamma_z&0\\
0&0&\gamma_x+\gamma_y
\end{pmatrix}.
\end{equation}

The Bloch-vector evolution equation~(\ref{eq:Blochevolution}) can be solved formally by defining
$\boldsymbol{T}(t)=R_{\boldsymbol{n}}^{-1}(\omega t)\boldsymbol{S}(t)$, which evolves according to
$d\boldsymbol{T}/dt=-4\Gamma_{\rm int}(t)\boldsymbol{T}$, where \begin{equation}
\Gamma_{\mathrm{int}}(t)=R_{\boldsymbol{n}}^{-1}(\omega t)\Gamma R_{\boldsymbol{n}}(\omega t).
\end{equation}
The transformation to $\boldsymbol{T}$ is equivalent to going to an interaction picture relative to the Hamiltonian $\half\omega\boldsymbol{n}\cdot\boldsymbol{\sigma}$.  A formal solution for $\boldsymbol{T}(\tau)$ can be written in terms of a time-ordered exponential; when translated back to the Bloch vector, the solution takes the form $\boldsymbol{S}(\tau)=B(\tau)\boldsymbol{S}(0)$, where
\begin{equation}
B(\tau)
=R_{\boldsymbol{n}}(\omega\tau)\!
\left[\,\mathbb{T}\,\exp\!\left(-4\int_0^\tau dt\,\Gamma_{\mathrm{int}}(t)\right)\right],
\end{equation}
with $\mathbb{T}$ denoting time ordering.  Inserted into Eqs.~(\ref{eq:ChoiQPxyz}), $B(\tau)$ gives the single-qubit error probabilities.

\section{Syndrome probabilities for the five-qubit code}\label{app:messy}
Consider an effective single-qubit channel of the form~(\ref{eq:Pxyzchannel}).  The syndrome probabilities (likelihood functions) for the five-qubit code experiencing the channel $\sF=\sG^{\otimes 5}$ are
\begin{subequations}
\begin{align}
 \prob(&S_0|P_x,P_y,P_z)\nonumber\\
&=\left(1-P_x+P_z\right) \Big(40 P_x^2 P_y P_z+20 P_x P_y^2 P_z-30 P_x P_y P_z+60 P_x^3 P_y+50 P_x^2 P_y^2-80 P_x^2 P_y+20 P_x P_y^3 \nonumber\\
&\qquad-45 P_x P_y^2+35 P_x P_y+11 P_x^3 P_z+P_x^2 P_z^2-18 P_x^2 P_z+P_x P_z^3-2 P_x P_z^2+8 P_x P_z+31 P_x^4-49 P_x^3+31 P_x^2\nonumber\\
&\qquad-9 P_x-5 P_y^2 P_z+5 P_y P_z+5 P_y^4-10 P_y^3+10 P_y^2-5 P_y+P_z^4-P_z^3+P_z^2-P_z+1\Big)\\
\prob(&S_1|P_x,P_y,P_z)\nonumber\\
&=\left(1-P_x+P_z\right) \Big(-P_x^2 P_y P_z+3 P_x P_y P_z^2-2 P_x P_y^2 P_z-15 P_x^3 P_y+P_x^2 P_y^2+11 P_x^2 P_y+4 P_x P_y^3-3 P_x P_y^2\nonumber\\
&\qquad -2 P_x P_y+2 P_x^3 P_z+5 P_x^2 P_z^2-3 P_x^2 P_z-4 P_x P_z^2+P_x P_z-14 P_x^4+17 P_x^3-7 P_x^2+P_x+P_y P_z^3-P_y^2 P_z^2-P_y P_z^2\nonumber\\
&\qquad+P_y^2 P_z+P_y^4-2 P_y^3+P_y^2+P_z^2\Big)\\
\prob(&S_2|P_x,P_y,P_z)\nonumber\\
&=\left(1-P_x+P_z\right) \Big(-7 P_x^2 P_y P_z-7 P_x P_y P_z^2-2 P_x P_y^2 P_z+8 P_x P_y P_z+3 P_x^3 P_y+P_x^2 P_y^2-P_x^2 P_y+4 P_x P_y^3\nonumber\\
&\qquad-3 P_x P_y^2-4 P_x^3 P_z-5 P_x^2 P_z^2+8 P_x^2 P_z-2 P_x P_z^3+5 P_x P_z^2-5 P_x P_z+4 P_x^4-4 P_x^3+P_x^2-P_y P_z^3-P_y^2 P_z^2\nonumber\\
&\qquad+3 P_y P_z^2+P_y^2 P_z-2 P_y P_z+P_y^4-2 P_y^3+P_y^2+P_z^3-P_z^2+P_z\Big)\\
\prob(&S_3|P_x,P_y,P_z)\nonumber\\
&=\left(1-P_x+P_z\right) \Big(4 P_x P_y P_z^2-2 P_x P_y P_z-12 P_x^2 P_y^2+6 P_x^2 P_y-12 P_x P_y^3+15 P_x P_y^2-5 P_x P_y-P_x^3 P_z\nonumber\\
&\qquad +P_x^2 P_z^2+P_x^2 P_z+P_x P_z^3-3 P_x P_z^2+4 P_x^4-4 P_x^3+P_x^2+2 P_y^2 P_z^2-2 P_y P_z^2-P_y^2 P_z+P_y P_z-3 P_y^4+6 P_y^3\nonumber\\
&\qquad-4 P_y^2+P_y+P_z^2\Big),
\end{align}
\end{subequations}
where the syndrome sets are
$S_0=\{0000\}$, $S_1=\{0001,0011,0110,1000,1100\}$, $S_2=\{0010,0100,0101,1001,1010\}$, and $S_3=\{0111,1011,1101,1110,1111\}$. Isotropic depolarizing noise is a special case with $\{P_x,P_y,P_z \}\mapsto \{P/3, P/3, P/3 \}$ the likelihood functions become
\begin{align}
\prob(0000|p)&=   1  -  \frac{5}{27}  \left(27 P  -54 P^2  +48 P^3 -16 P^4\right) \\
\prob(S|p)&=\frac{1}{81} \left(27 P  -54 P^2  +48 P^3 -16 P^4\right)
\end{align}
where the syndrome set $S$ contains all syndromes except the trivial one, i.e., $S\in \{ 0001, ...., 1111\}$.
\end{widetext}


\begin{thebibliography}{99}

\bibitem{Shor95}
P.~W.~Shor, {\em Scheme for reducing decoherence in quantum computer memory},
\href{http://dx.doi.org/10.1103/PhysRevA.52.R2493}{Physical Review~A {\bf 52}, 2493 (1995)}.

\bibitem{Steane96}
A.~Steane, {\em Multiple-particle interference and quantum error correction},
\href{http://dx.doi.org/10.1098/rspa.1996.0136}{Proceedings of the Royal Society of London~A
{\bf 452}, 2551 (1996)}.

\bibitem{Got96}
D.~Gottesman, {\em A class of quantum error-correcting codes saturating the quantum Hamming bound},
\href{http://dx.doi.org/10.1103/PhysRevA.54.1862}{Physical Review~A {\bf 54}, 1862 (1996)}.

\bibitem{Kitaev97}
A.~Yu. Kitaev, {\em Fault-tolerant quantum computation by anyons},
\href{http://dx.doi.org/10.1016/S0003-4916(02)00018-0}{Annals of Physics {\bf 303}, 2 (2003)}; see also \href{http://arxiv.org/abs/quant-ph/9707021}{quant-ph/9707021}.
	


\bibitem{MCMC}
M.~S. Arulampalam, S.~Maskell, N.~Gordon, and T.~Clapp, {\em A tutorial on particle filters for online nonlinear/non-Gaussian Bayesian tracking}, \href{http://dx.doi.org/10.1109/78.978374}{IEEE Transactions on Signal Processing {\bf 50}, 174 (2002)}; A. Doucet and A. M. Johansen, {\em A tutorial on particle filtering and smoothing: Fifteen years later}, \href{http://automatica.dei.unipd.it/tl_files/utenti/lucaschenato/Classes/PSC10_11/Tutorial_PF_doucet_johansen.pdf}{The Oxford Handbook of Nonlinear Filtering, Vol.~12 (Oxford University Press, 2009), p.~656.}

\bibitem{BriDelas12}
H.~J. Briegel and G. De las Cuevas, {\em Projective simulation for artificial intelligence},
\href{http://dx.doi.org/10.1038/srep00400}{Scientific Reports {\bf 2}, 400 (2012).}

\bibitem{LafMiqPazZur96}
R.~Laflamme, C.~Miquel, J.~P. Paz, and W.~H.~Zurek, {\em Perfect quantum error correcting code},
\href{http://dx.doi.org/10.1103/77.198}{Physical Review Letters {\bf 77}, 198 (1996).}

\bibitem{BenDiVSmoWoo96}
C.~H. Bennett, D.~P. DiVincenzo, J.~A. Smolin, and W.~K. Wootters, {\em Mixed-state entanglement and quantum error correction},
\href{http://dx.doi.org/10.1103/PhysRevA.54.3824}{Physical Review~A {\bf 54}, 3824 (1996)}.

\bibitem{DurSkoFroKra13}
W. D\"ur, M. Skotiniotis, F. Fr\"owis, and B. Kraus, {\em Improved quantum metrology using quantum error-correction},  \href{http://dx.doi.org/10.1103/PhysRevLett.112.080801}{Physical Review Letters {\bf 112}, 080801 (2014)}.

\bibitem{Ozeri13}
R. Ozeri, {\em Heisenberg limited metrology using quantum error-correction codes}, \href{http://arxiv.org/abs/1310.3432v1}{arXiv:1310.3432v1}.

\bibitem{ArrVinAhaRet13}
G. Arrad, Y. Vinkler, D. Aharonov, and A. Retzker, {\em Increasing sensing resolution with error correction}, \href{http://dx.doi.org/10.1103/PhysRevLett.112.150801}{Physical Review Letters { \bf 112}, 150801 (2014)}.

\bibitem{KesLovSusLuk13}
E. M. Kessler, I. Lovchinsky, A. O. Sushkov, and M. D. Lukin, {\em Quantum error correction for metrology}, \href{http://dx.doi.org/10.1103/PhysRevLett.112.150802}{Physical Review Letters { \bf 112}, 150802 (2014)}.


\bibitem{FowSankKelBarMar2014}
A. G. Fowler, D. Sank, J. Kelly, R. Barends, and J. M. Martinis, {\em Scalable extraction of error models from the output of error detection circuits}, \href{http://arxiv.org/abs/1405.1454}{arXiv:1405:1454}.

\bibitem{OmSriBan2014}
S. Omkar, R. Srikanth, and S. Banerjee, {\em Online characterization of quantum dynamics using quantum error correction}, \href{http://arxiv.org/abs/1405.0964}{arXiv:1405.0964}.

\bibitem{MikeandIke}
M.~A. Nielsen and I.~L. Chuang, {\em Quantum Computation and Quantum Information\/} (Cambridge University Press, 2000).

\bibitem{pnote}
The probability $p(\tau)$ can be simply obtained by assuming that a qubit has probability $2\gamma\,dt$ to bit-flip in the time interval $dt$; thus $p$ satisfies the differential equation $dp/dt=2\gamma(1-2p)$, whose solution for a time interval $\tau$ is Eq.~(\ref{p_timedepend}).   This approach to the master equation~(\ref{eq:vacuum_master}) is worked out in G.~J. Milburn, {\em Quantum control based on measurement}, in {\em Quantum Information and Coherence}, edited by E.~Andersson and P.~{\"O}hberg (Springer, Berlin, 2014), pp.~147--158.

\bibitem{QInfer}
C.~Granade and C.~Ferrie, \href{https://github.com/csferrie/python-qinfer}{{{QInfer}:~Library for Statistical Inference in Quantum Information}, (2012--)}.

\bibitem{WieGraFer13}
N.~Wiebe, C.~E. Granade, C.~Ferrie, D.~G. Cory, {\em Quantum Hamiltonian learning using imperfect quantum resources }, \href{http://arxiv.org/abs/1309.0876}{Physical Review~A {\bf 89}, 042314 (2014)}.

\bibitem{SerChaCom11}
A.~Sergeevich, A.~Chandran, J.~Combes, S.~D. Bartlett, H.~M. Wiseman, {\em Characterization of a qubit Hamiltonian using adaptive measurements in a fixed basis},
\href{http://dx.doi.org/10.1103/PhysRevA.84.052315}{Physical Review~A {\bf 84}, 052315 (2011)}.

\bibitem{FerGraCor13}
C.~Ferrie, C.~Granade, and D.~G.~Cory, {\em How to best sample a periodic probability distribution, or on the accuracy of Hamiltonian finding strategies},
\href{http://dx.doi.org/10.1007/s11128-012-0407-6}{Quantum Information Processing {\bf 12}, 611 (2013)}.


\bibitem{RahDohMab02}
B.~Rahn, A. C. Doherty, and H.~Mabuchi, Hideo, {\em Exact performance of concatenated quantum codes}, \href{10.1103/PhysRevA.66.032304}{Physical Review~A {\bf 66}, 032304 (2002)}.

\bibitem{chernoff1952}
H.~Chernoff, {\em A measure of asymptotic efficiency for tests of a hypothesis based on the sum of observations},
\href{http://dx.doi.org/10.1214/aoms/1177729330}{The Annals of Mathematical Statistics {\bf 23}, 493 (1952)}.

\bibitem{HoeffdingChernoff1963}
W.~Hoeffding, {\em Probability inequalities for sums of bounded random variables},
\href{http://dx.doi.org/10.1080/01621459.1963.10500830}{Journal of the American Statistical Association {\bf 58}, (1963)}.

\bibitem{SchvanEnk13}
L. Schwarz and S.~J.~van Enk, {\em Error models in quantum computation: An application of model selection}, \href{http://link.aps.org/doi/10.1103/PhysRevA.88.032318}{Phys. Rev. A {\bf 88}, 032318 (2013)}.

\bibitem{Preskill}
J. Preskill, {\em Quantum clock synchronization and quantum error correction}, \href{http://arxiv.org/abs/quant-ph/0010098v1}{quant-ph/0010098v1}.

\bibitem{MerGamSmo13}
S.~T. Merkel, J.~M. Gambetta, J.~A. Smolin, S.~Poletto, A.~D. C\'orcoles, B.~R. Johnson, C.~A. Ryan, and M.~Steffen, {\em Self-consistent quantum process tomography}, \href{http://dx.doi.org/10.1103/PhysRevA.87.062119}{Physical Review~A {\bf 87}, 062119 (2013)}.

\bibitem{FlaLui11}
S.~T. Flammia and Y.-K. Liu, {\em Direct fidelity estimation from few Pauli measurements}, \href{http://dx.doi.org/10.1103/PhysRevLett.106.230501}{Physical Review Letters {\bf 106}, 230501 (2011)}.

\bibitem{daSlLanPou11}
M.~P. da Silva, O.~Landon-Cardinal, and D.~Poulin, {\em Practical characterization of quantum devices without tomography}, \href{http://dx.doi.org/10.1103/PhysRevLett.107.210404}{Physical Review Letters {\bf 107}, 210404 (2011)}.

\bibitem{KimdaSRya14}
S. Kimmel, M.~P. da Silva, C.~A. Ryan, B.~R. Johnson, and T.~Ohki, {\em Robust extraction of tomographic information via randomized benchmarking}, \href{http://dx.doi.org/10.1103/PhysRevX.4.011050}{Physical Review~X {\bf 4}, 011050 (2014)}.

\bibitem{Choi75}
M.-D.~Choi, {\em Completely positive linear maps on complex matrices},
\href{http://dx.doi.org/10.1016/0024-3795(75)90075-0}
{Linear Algebra and Its Applications {\bf 10}, 285 (1975)}.










\end{thebibliography}
\end{document}